\newcommand{\TypeA}{${[a]}$}
\newcommand{\TypeAtext}{Radial velocity maximum}
\newcommand{\TypeB}{${[b]}$}
\newcommand{\TypeBtext}{Periastron}
\newcommand{\TypeC}{${[c]}$}
\newcommand{\TypeCtext}{Cool in front}
\newcommand{\TypeD}{${[d]}$}
\newcommand{\TypeDtext}{Primary in front}
\newcommand{\TypeE}{${[e]}$}
\newcommand{\TypeEtext}{Primary behind}
\newcommand{\flip}{``flip--flop''}
\newcommand{\genflip}{``stationary flip--flop''}
\documentclass[iop]{emulateapj}     
\newcommand{\tablepage}{}           
\shorttitle{General Model for
Light curves of chromospherically active binary stars}
\shortauthors{Jetsu et al.}
\begin{document}
\title{General Model for Light curves \\
of Chromospherically Active Binary Stars}
\author{L. Jetsu}  
\affil{Department of Physics,
P.O. Box 64, FI-00014 University of Helsinki, Finland}
\email{lauri.jetsu@helsinki.fi}
\author{G.W. Henry}
\affil{Center of Excellence in Information Systems, Tennessee State University, 
3500 John A. Merritt Blvd., Box 9501, Nashville, TN 37209, USA}
\author{J. Lehtinen}
\affil{Max-Planck-Institut f\"ur Sonnensystemforschung,
Justus-von-Liebig-Weg 3, 37077 G\"ottingen, Germany}
\begin{abstract}
The starspots on the surface of many chromospherically active binary
stars concentrate on long--lived active longitudes separated by 180 degrees.
The activity shifts between these two longitudes, the \flip ~events, 
have been observed
in single stars like FK Comae and binary stars like $\sigma$ Geminorum.
Recently, interferometry has revealed that ellipticity may at least partly
explain the \flip ~events in $\sigma$ Geminorum. 
This idea was supported by
the double peaked shape of the long--term mean light curve of this star.
Here, we show that the long--term mean light curves 
of fourteen chromospherically
active binaries follow a general model which
explains the connection between
orbital motion, starspot distribution changes, ellipticity
and \flip ~events.
Surface differential rotation is probably weak in these stars,
because the interference of two constant period waves may
explain the observed light curve changes.
These two constant periods are the active longitude period $(P_{\mathrm{act}})$
and the orbital period $(P_{\mathrm{orb}})$.
We also show how to apply the same model to single stars,
where only the value of $P_{\mathrm{act}}$ is known.
Finally, we present a tentative interference hypothesis about 
the origin of magnetic fields in all spectral types of stars.
\end{abstract}

\section{Introduction} \label{introduction}

An ancient Egyptian calendar of 
lucky and unlucky days, the Cairo Calendar,
is the oldest preserved historical document 
of the discovery of a variable star,
\object{Algol} \citep{Por08,Jet13,Jet15}.
Today, the General Catalogue of Variable Stars 
contains nearly 50,000 stars.
There are numerous classes of variable stars and
the classification criteria are constantly updated 
\citep[][]{Sam97}.
Different classes have their
own typical light curves \citep{Dra14}.
One class of variable stars
is called the Chromospherically Active Binary Stars 
(hereafter CABS).
The third catalogue of CABS lists 409
such binaries \citep{Eke08}.

In his review of starspots,
\cite{Str09} wrote that the ``starspot hypothesis'' 
was first presented by a French astronomer Ismael Boulliau (1605--1694)
to explain the variability of Mira.
This star, also known as $o$ Ceti,
was the first variable star
discovered by ``modern'' astronomers
in 1596 (David Fabricius, 1564--1617).
Boulliau's hypothesis was unfortunately not true,
because pulsations cause the variability of Mira.
According to \cite{Str09},
the first observations of starspots were 
made by \cite{Kro47} in the light curves of four eclipsing binaries.
The presence of this phenomenon in the light
curves of active stars was firmly established later
by \cite{Hof65,Chu66,Cat67,Hal72}.

Unlike the longitudinally evenly distributed sunspots in the Sun,
the starspots in active single \citep[e.g.][FK Com]{Jet93}
or binary \citep[e.g.][$\sigma$ Gem]{Jet96} stars
concentrate on long--lived active longitudes and seem to
undergo shifts of about 180 degrees in longitude.
The presence of this \flip ~phenomenon in the CABS,
$\sigma$ Gem, was questioned by \citet{Roe15}.
Their interferometric observations revealed that ellipticity
of $\sigma$ Gem may explain the stability of the two minima
in the long--term mean light curve (hereafter MLC).
Recently, \citet{Sil16} reported that another CABS, 
BM CVn, has a sinusoidal MLC with an amplitude of $0.^{\mathrm{m}}042$.
Ellipticity fails to explain the MLC shape of BM CVn.
Here, we study the light curves of fourteen CABS. 
Our sample includes the light curves of 
the two above mentioned CABS, $\sigma$ Gem and BM CVn.

We use the following abbreviations

\begin{itemize}

\item[] CABS = Chromospherically Active Binary Star 

\item[] MLC = Mean Light Curve 

\item[] CPS = Continuous Period Search

\item[] A  = More active CABS member

\item[] B  = Other CABS member

\item[] Aa = A in front of B epoch

\item[] Ac = B in front of A epoch

\item[] Ab = Mid epoch between Aa and Ac

\item[] Ad = Mid epoch between Ac and Aa

\item[] S1 = Larger starspot on A 

\item[] S2 = Smaller starspot on A 

\item[] S1f = S1 visibility maximum at Aa \& No S2     

\item[] S1b = S1 visibility maximum  at Ac \& No S2    

\item[] S12fb = S1 visibility maximum \& S2 unseen at Aa 

\item[] S12bf = S1 unseen \& S2 visibility maximum at Aa 

\end{itemize}

The meaning of the last ten abbreviations is explained
in greater detail in Sect. \ref{stationary}.

\section{Observations}

Our differential standard Johnson $V$ 
photometric observations
were made with the T3 0.4m automated
photoelectric telescope (APT) at Fairborn Observatory in Arizona.
The information of our photometry 
of fourteen CABS is summarized in the seven first columns of 
Table 
\ref{resultsone}.
The accuracy of the differential $V$
magnitudes is approximately $0.^{\mathrm{m}}006$.
\cite{Hen99} and \cite{Fek05} have described 
the operation of the T3 0.4~m APT
and photometric data reduction procedures
in greater detail.

\section{MLC analysis \label{mlcanalysis}}

The original data are the differential $V$ magnitude observations
$m_i=m(t_i)$, where $t_i$ is the observing time.
The time points are transformed into phases
$\phi_i={\mathrm{FRAC}}[(t_i-t_0)/P_{\mathrm{orb}}]$,
where ${\mathrm{FRAC}}$ removes the integral
part of its argument, 
i.e. the number of full orbital rounds $P_{\mathrm{orb}}$  
completed after $t_0$.
The $P_{\mathrm{orb}}$ and $t_0$ values
of each CABS \citep{Eke08}
are given in Table \ref{resultsone}.
The  $m_i$ observations are binned in phase,
using $N=20$ evenly spaced bins,
where the limits of the $j$:th bin are
$(j-1)/N \le \phi_i \ < j/N$.
A bin must contain at least two $m_i$ values.
The binned data for the $n_j$ values
of $m_i$ in the $j$:th bin are
$x_j = (j/N)-0.25$,
$y_j = n_j^{-1} \sum_{i=1}^{n_j} m_i$,
$\sigma_j  =  n_j^{-1/2} [ n^{-1}_j \sum_{i=1}^{n_j} (m_i-y_j)^2]^{1/2} 
               = n_j^{-1} [\sum_{i=1}^{n_j} (m_i-y_j)^2]^{1/2}$.

The model for these data is
\begin{eqnarray}
g(x,\bar{\beta}) = a_0 + \sum_{k=1}^K a_i \cos (k x) + b_i \sin(k x),
\label{MLCmodel}
\end{eqnarray}
where $\bar{\beta}=[a_0,a_1, ..., a_K,b_1,...,b_K]$ 
are the free parameters.
Note that the model is simply $g(t,\bar{\beta})=a_0$, if $K=0$.
The model residuals $\epsilon_j= y_j - g(x_j,\bar{\beta})$ give
$\chi^2(\bar{y},\bar{\beta})
=\sum_{j=1}^{N} w_j \epsilon_j$, where $w_j=\sigma_j^{-2}$.

The main question is, do these data contain a periodic signal?
And if so, what is the correct order $K$ for the model of this signal?
We solve this problem, 
as \cite{Leh11} did,
by computing the Bayesian information criterion parameter
\begin{eqnarray}
R_{\mathrm{BIC}}= 2 n ~ {\mathrm{ln}} \lambda (\bar{y},\bar{\beta})
              + (5K+1) ~ {\mathrm{ln}} ~n,
\label{BIC}
\end{eqnarray}
where $\lambda (\bar{y},\bar{\beta}) = 
\chi^2 (\bar{y}, \bar{\beta}) [\sum_{j=1}^{N}w_j]^{–1}$.
The best modelling order $K$ for the data is the 
value of the order that minimizes $R_{\mathrm{BIC}}$.
We test the values $K=0, 1$ and 2
for the binned data of each CABS. 
The best $K$ value for the MLC of all data of each star,
and the peak to peak amplitude $A_{\mathrm{All}}$ of this MLC, 
are given in Table \ref{resultsone}.
The periodic
MLC phenomenon is present in all fourteen CABS (i.e. $K \ge 1$).

\begin{figure} 
\resizebox{\hsize}{!}{\includegraphics{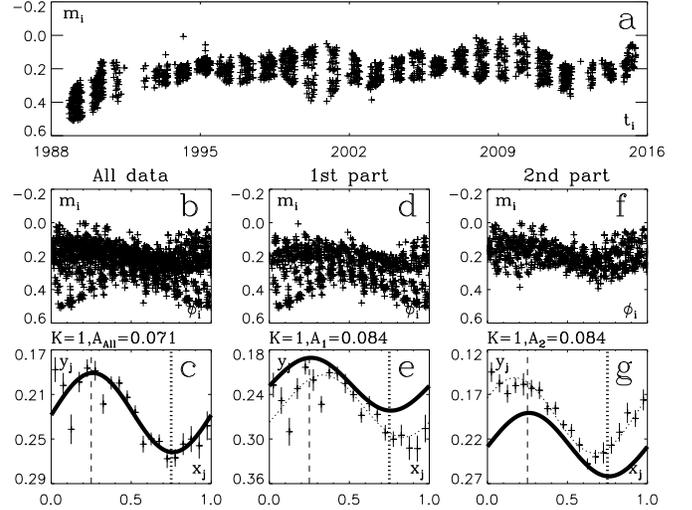}}
\caption{\label{dmuma} DM~UMa,
{\bf (a)} All data: $V$ magnitudes (crosses) versus time.
{\bf (b)} All data: $V$ magnitudes (crosses) versus orbital phase.
{\bf (c)} All data: Binned magnitudes (crosses with error bars), 
MLC (thick continuous line), order ($K$) and amplitude ($A_{\mathrm{All}}$).
Epoch Aa (thick dotted vertical line) and epoch Ac (thin dashed vertical line).
{\bf (d)} 1st part of data:  $V$ magnitudes (crosses) versus orbital phase. 
{\bf (e)} 1st part of data: Binned magnitudes (crosses with error bars), 
MLC (thin dashed line), order ($K$), amplitude ($A_1$) and MLC 
of all data (thick line from ``a'').
Epoch Aa (thick dotted vertical line) and epoch Ac (thin dashed vertical line).
{\bf (fg)} 2nd part of data: otherwise as in ``de'' }
\end{figure}

We divide the data in two parts in our Figs. \ref{dmuma} -- \ref{iipeg}.
This shows, if MLC changes.
The 1st and 2nd parts are before and after $t_1+\Delta T/2$, 
respectively.
The order $K$ of the MLC model for the
1st and 2nd part of data is fixed to the $K$ value
obtained for all data.
However, the amplitudes of these MLC models, 
$A_1$ and $A_2$, are determined separately 
from a fit to the binned data of the 1st and 2nd part of data.
We use the notation $\Delta A$ for the maximum difference 
between the three amplitudes
$A_{\mathrm{All}}$, $A_1$ and $A_2$.

The orbital ephemeris epoch ($t_0$), the orbital period ($P_0$),
the eccentricity ($e$) and the spectral type of 
A member of CABS
information in Tables \ref{resultsone} and \ref{resultstwo} is
from \citet[][third catalogue of CABS]{Eke08}.
The original references for this information are given
in Sect. \ref{results}, where the results
for each individual star are discussed separately.
We use the radial velocity maximum epochs of Table \ref{resultstwo}
(Ab epochs) to compute the orbital phases from
\begin{eqnarray}
\phi_{\mathrm{orb}}={\mathrm{FRAC}}[(t-t_0)/P_{\mathrm{orb}}],
\label{thomasinvaiheet}
\end{eqnarray}
where $t_0$ is the Ab epoch given
in Table \ref{resultstwo}.

\tablepage
\begin{table*}
\begin{center}
\caption{\label{resultsone} CABS sample.}
\addtolength{\tabcolsep}{-0.04cm}
\begin{tabular}{lllllcclllcc}
\tableline \tableline
Variable       &       & Comparison        & Beg  & End   & 
$\Delta T$     & $n~~$ & ~~$P_{\mathrm{orb}}$ & $t_0$ & $e$ & $K$ & $~~A_{\mathrm{All}}$ \\
                &       &                   &      &       &
$[{\mathrm{y}}]$ &       & $[{\mathrm{d}}]$  & $[{\mathrm{HJD-2400000}}]$ & & & $[{\mathrm{mag}}]$ \\
\tableline
              DM~UMa &  SAO~15334 &   HD~95362 &      29.10.1988 &       13.6.2015  &  26.6 &  1915 &           7.492 &         43881.4    \TypeA &            0 &  1 &  0.071 \\
              XX~Tri &   HD~12545 &   HD~12478 &       7.10.1990 &       13.2.2015  &  24.4 &  1962 &        23.96924 &       47814.315    \TypeA &            0 &  1 &  0.136 \\
              EL~Eri &   HD~19754 &   HD~19421 &      19.10.1990 &       11.2.2015  &  24.3 &  1642 &          48.263 &         44419.9    \TypeB &          0.1 &  1 &  0.048 \\
            V711~Tau &   HD~22468 &   HD~22484 &      13.11.1987 &       11.3.2015  &  27.3 &  2293 &         2.83774 &       51142.943    \TypeC &            0 &  2 &  0.030 \\
              EI~Eri &   HD~26337 &   HD~26409 &       11.1.1988 &        7.3.2015  &  27.2 &  2268 &        1.947227 &       46091.052    \TypeA &            0 &  2 &  0.017 \\
           V1149~Ori &   HD~37824 &   HD~38309 &       13.9.1988 &       28.3.2015  &  26.5 &  2274 &        53.57465 &       48625.022    \TypeA &            0 &  1 &  0.050 \\
        $\sigma$~Gem &   HD~62044 &   HD~60318 &      21.11.1987 &        1.5.2015  &  27.4 &  2984 &        19.60447 &        47227.15    \TypeD &        0.012 &  2 &  0.039 \\
              FG~Uma &   HD~89546 &   HD~90400 &       15.3.1992 &       11.6.2015  &  23.2 &  3145 &        21.35957 &       49297.702    \TypeA &            0 &  2 &  0.045 \\
              HU~Vir &  HD~106225 &  HD~105796 &        6.4.1990 &       15.6.2015  &  25.2 &  2062 &       10.387552 &       49993.195    \TypeB &       0.0093 &  2 &  0.136 \\
              BM~CVn &  HD~116204 &  HD~116010 &        6.4.1990 &       17.6.2015  &  25.2 &  2917 &         20.6252 &        45251.62    \TypeA &            0 &  1 &  0.040 \\
            V478~Lyr &  HD~178450 &  HD~177878 &      14.11.1987 &       21.6.2015  &  27.6 &  2641 &       2.1305140 &       45939.801    \TypeA &            0 &  2 &  0.010 \\
           V1762~Cyg &  HD~179094 &   HD177483 &       25.4.1988 &       21.6.2015  &  27.2 &  2526 &        28.58973 &       31043.408    \TypeA &            0 &  2 &  0.031 \\
              HK~Lac &  HD~209813 &  HD~210731 &       30.4.1988 &       21.6.2015  &  27.1 &  2454 &         24.4284 &        40017.17    \TypeB &         0.01 &  1 &  0.089 \\
              II~Peg &  HD~224085 &  HD~224930 &      16.11.1987 &       23.1.2015  &  27.2 &  2049 &        6.724333 &      49582.9268    \TypeE &            0 &  1 &  0.064 \\
\end{tabular}
\tablecomments{
Variable (Cols 1-2: Variable designation and HD or SAO number), 
comparison star (Col 3: SAO- or HD-number), 
first and last observing date (Cols 4-5: Beg and End),
time span and number of observations 
(Cols 6-7: $[\Delta T]={\mathrm{y}}$ and $n$),
orbital period, epoch and eccentricity 
(Cols 8-10: $P_{\mathrm{orb}}={\mathrm{d}}$, $t_0$ and $e$, 
epoch types: \TypeA = \TypeAtext, \TypeB = \TypeBtext, 
\TypeC = \TypeCtext, \TypeD = \TypeDtext, \TypeE = \TypeEtext),
MLC order and amplitude (Cols 11-12: K and $[A_{\mathrm{All}}]={\mathrm{mag}}$). }
\addtolength{\tabcolsep}{+0.04cm}
\end{center}
\end{table*}

\tablepage
\begin{table}
\begin{center}
\caption{\label{resultstwo} CABS sample.}
\addtolength{\tabcolsep}{-0.08cm}
\begin{tabular}{llllll}
\tableline \tableline
Variable   &     & Sp-type of A & Aa  & Ab  & Ac  \\
\tableline
         DM~UMa &    SB1 &   K0--1~IV--III &  43879.527 &  43881.400 &  43883.273 \\
         XX~Tri &    SB1 &          K0~III &  47808.323 &  47814.315 &  47820.307 \\
         EL~Eri &    SB1 &      G8~III--IV &  44419.966 &  44432.032 &  44444.097 \\
       V711~Tau &    SB2 &           K1~IV &  51142.943 &  51143.652 &  51144.362 \\
         EI~Eri &    SB1 &           G5~IV &  46090.565 &  46091.052 &  46091.539 \\
      V1149~Ori &    SB1 &          K0~III &  48611.628 &  48625.022 &  48638.416 \\
   $\sigma$~Gem &    SB1 &          K1~III &  47227.150 &  47232.051 &  47236.952 \\
         FG~Uma &    SB1 &          G9~III &  49292.362 &  49297.702 &  49303.042 \\
         HU~Vir &    SB1 &          K2~III &  49993.403 &  49996.000 &  49998.597 \\
         BM~CVn &    SB1 &          G8~III &  45246.464 &  45251.620 &  45256.776 \\
       V478~Lyr &    SB1 &            G8~V &  45939.268 &  45939.801 &  45940.334 \\
      V1762~Cyg &    SB2 &      K2~IV--III &  31036.261 &  31043.408 &  31050.555 \\
         HK~Lac &    SB1 &          K0~III &  40011.066 &  40017.173 &  40023.281 \\
         II~Peg &    SB1 &            K2~V &  49579.565 &  49581.246 &  49582.927 \\
\end{tabular}
\addtolength{\tabcolsep}{+0.08cm}
\tablecomments{Variable (Col. 1), Single-lined (SB1) or double-lined (SB2),
Sp-type of member A, epochs Aa, Ab and Ac [HJD-2400000] 
(Cols 2-6)}
\end{center}
\end{table}


\section{MLC analysis results} \label{results}

Here, we describe the MLC of each individual CABS and give
the original references for their physical parameters in Tables
 \ref{resultsone} and \ref{resultstwo}.

\subsection{MLC of DM UMa}

The high amplitude, $A_{\mathrm{All}}=0.^{\mathrm{m}}071$, sinusoidal MLC of
\object{DM~UMa} remains nearly unchanged between 
$0.4 < \phi_{\mathrm{orb}} < 0.6$ (Figs. \ref{dmuma}ceg).
The largest MLC changes occur between 
Aa $\equiv -0.25 < \phi_{\mathrm{orb}} < 0.25 \equiv$ Ac.
MLC level, shape and  phase are nearly stable.
MLC amplitude changes are small 
($\Delta A=0.^{\mathrm{m}}013$).
MLC minimum coincides with Aa
\citep[][$P_{\mathrm{orb}}$, $t_0$, $e$, Sp-type of member A]{Cra79,Gle95,Bar98,Hat98}.

\begin{figure} 
\resizebox{\hsize}{!}{\includegraphics{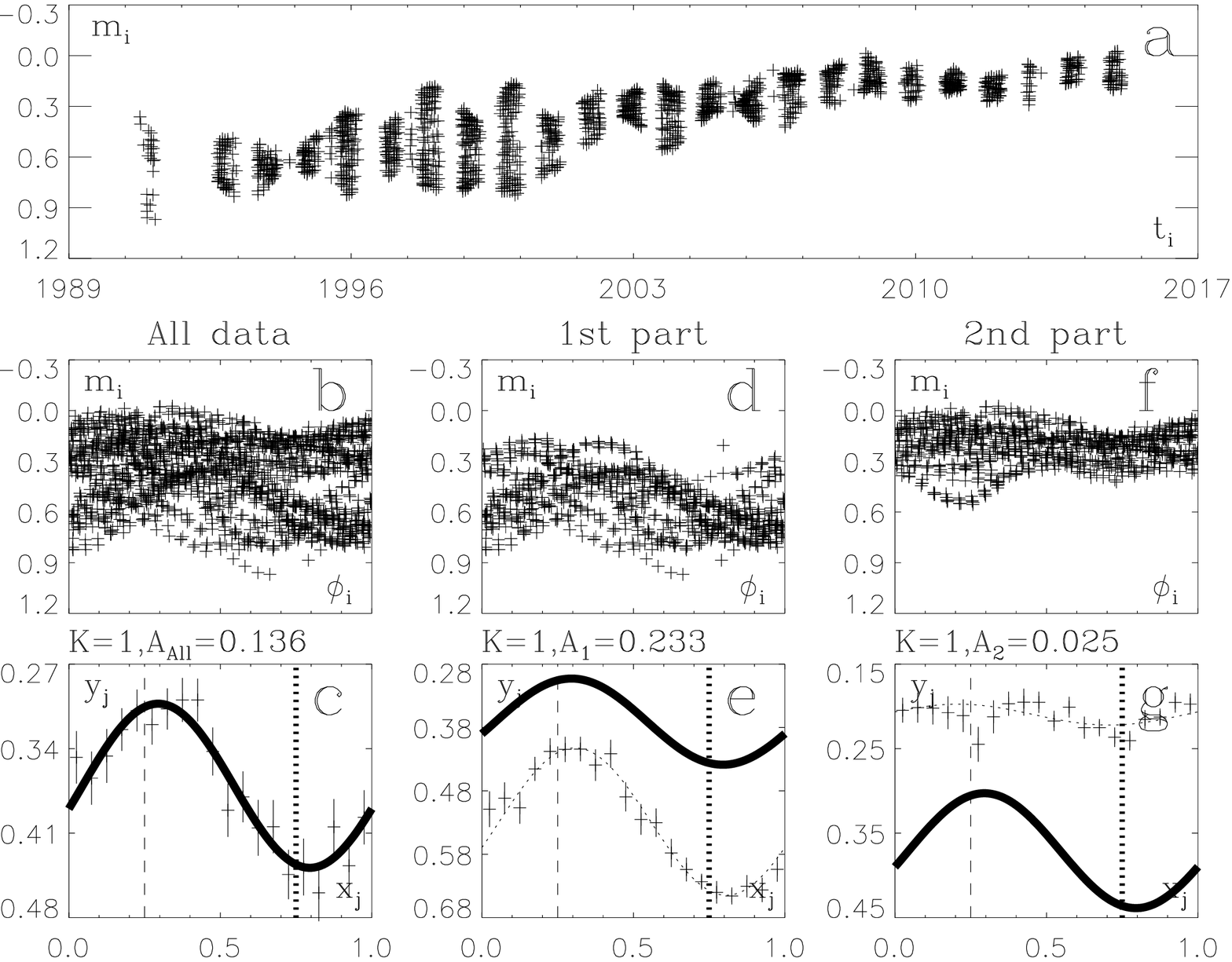}}
\caption{XX~Tri , otherwise as in Fig. \ref{dmuma}.
\label{xxtri}}
\end{figure}

\subsection{MLC of XX Tri}

The sinusoidal MLC of \object{XX~Tri} has an extremely high amplitude
of $A_{\mathrm{All}}=0.^{\mathrm{m}}136$. 
This variation increases to $A_1=0.^{\mathrm{m}}233$
during the 1st part of data, and then
decreases to $A_2=0.^{\mathrm{m}}025$ in the 2nd part.
Despite of these dramatic changes,
MLC phase remains stable.
MLC minimum and maximum phases 
coincide with Aa and Ac epochs 
\citep[][$P_{\mathrm{orb}}$, $t_0$, $e$, Sp-type of member A]{Bop93,Str92,Str99}.
Note the minor MLC dips at $\phi_{\mathrm{orb}}=0.25$ and 0.75 in Fig. \ref{xxtri}g
which will be discussed later in Sect. \ref{argumentfour}.

\begin{figure} 
\resizebox{\hsize}{!}{\includegraphics{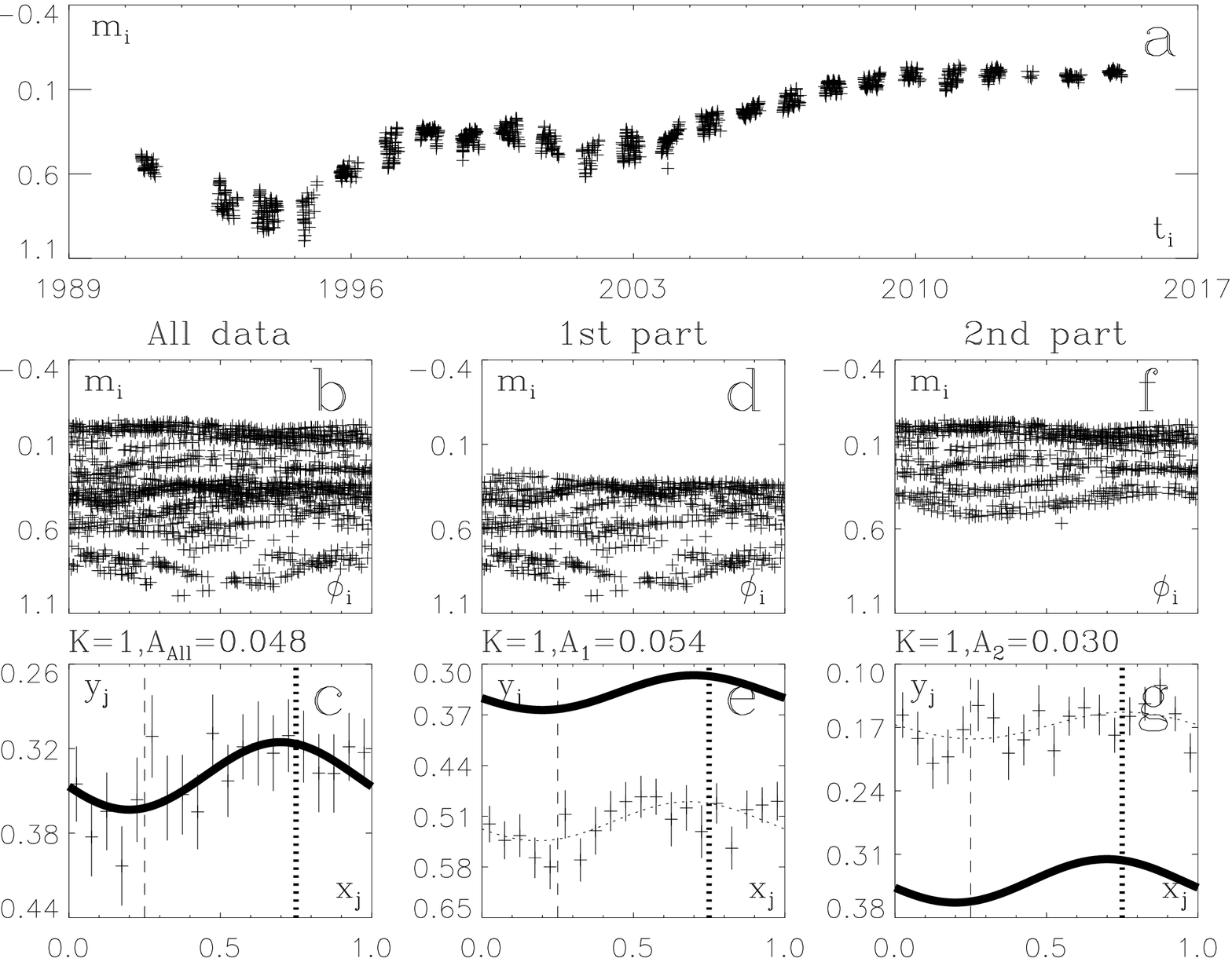}}
\caption{EL~Eri, otherwise as in Fig. \ref{dmuma}.
\label{eleri}}
\end{figure}

\subsection{MLC of EL Eri}


\object{EL~Eri} has a sinusoidal MLC with an amplitude
of $A_{\mathrm{All}}=0.^{\mathrm{m}}048$ 
(Fig. \ref{eleri}c). 
MLC shape, amplitude and phase remain quite stable 
($\Delta A=0.^{\mathrm{m}}018$),
despite of the large mean level changes between 1st and 2nd part
(Figs \ref{eleri}{c--g}).
The Aa and Ac epochs nearly coincide with MLC maximum and minimum
\citep[][$P_{\mathrm{orb}}$, $t_0$, $e$, Sp-type of member A]{Bal87,Fek86}.

\begin{figure} 
\resizebox{\hsize}{!}{\includegraphics{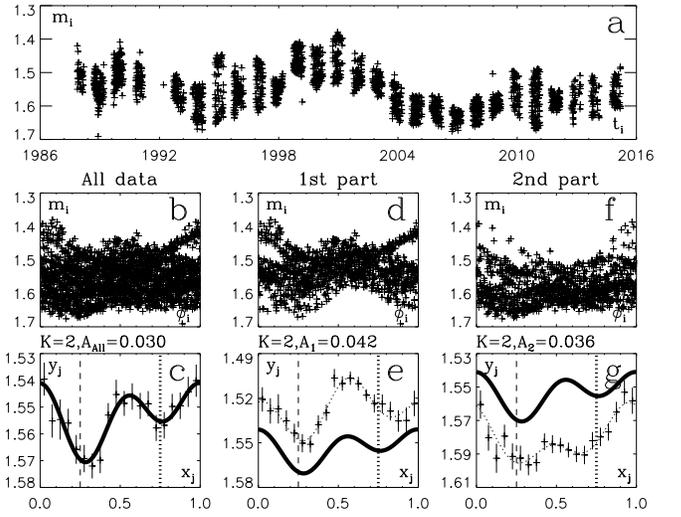}}
\caption{V711~Tau, otherwise as in Fig. \ref{dmuma}.
\label{v711tau}}
\end{figure}

\subsection{MLC of V711 Tau}

The double peaked MLC of \object{V711~Tau} has a low amplitude
(Fig. \ref{v711tau}: $A_{\mathrm{All}}=0.^{\mathrm{m}}030$).
This amplitude remains quite stable (Figs. \ref{v711tau}ceg:
$\Delta A=0.^{\mathrm{m}}012$), 
regardless of the changes in the mean level (Figs. \ref{v711tau}df).
MLC primary minimum phase $\phi_{\mathrm{orb}}=0.25$ does not shift,
but the secondary minimum $\phi_{\mathrm{orb}}=0.8$ in the 1st part
shifts to $0.7$ in the 2nd part.
The epochs of Ac and Aa coincide
with the MLC primary and secondary minima
\citep[][$P_{\mathrm{orb}}$, $t_0$, $e$, [Sp-type of member A]{Fek83,Don92,Gar03}.
However, the interpretation this MLC may be more complicated,
because the hotter member B (G5~IV)
could also influence MLC \citep{Fek83,Gar03}.

\begin{figure} 
\resizebox{\hsize}{!}{\includegraphics{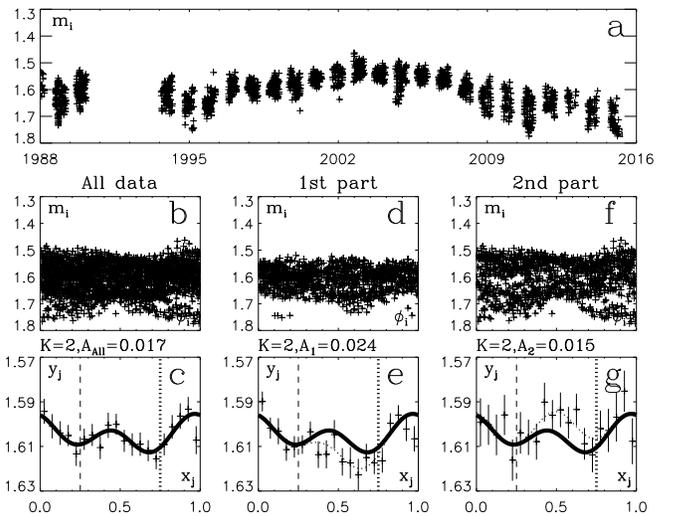}}
\caption{EI~Eri, otherwise as in Fig. \ref{dmuma}.
\label{eieri}}
\end{figure}

\subsection{MLC of EI Eri}

\object{EI~Eri} has a very stable low amplitude
double peaked MLC (Fig. \ref{eieri}a: $A_{\mathrm{All}}=0.^{\mathrm{m}}017$).
MLC mean and amplitude do not change.
Only minor MLC changes occur in the interval $0.25 < \phi_{\mathrm{orb}} < 0.75$.
MLC primary minimum at $\phi_{\mathrm{orb}}=0.65$ in the 1st part
shifts to $\phi_{\mathrm{orb}}=0.20$ in the 2nd part.
Epochs of Ac and Aa occur about $\Delta \phi_{\mathrm{orb}}=0.05$ after
MLC secondary and primary minima
\citep[][$P_{\mathrm{orb}}$, $t_0$, $e$, Sp-type of member A]{Fek86,Fek87,Str90,Cut95}.

\begin{figure} 
\resizebox{\hsize}{!}{\includegraphics{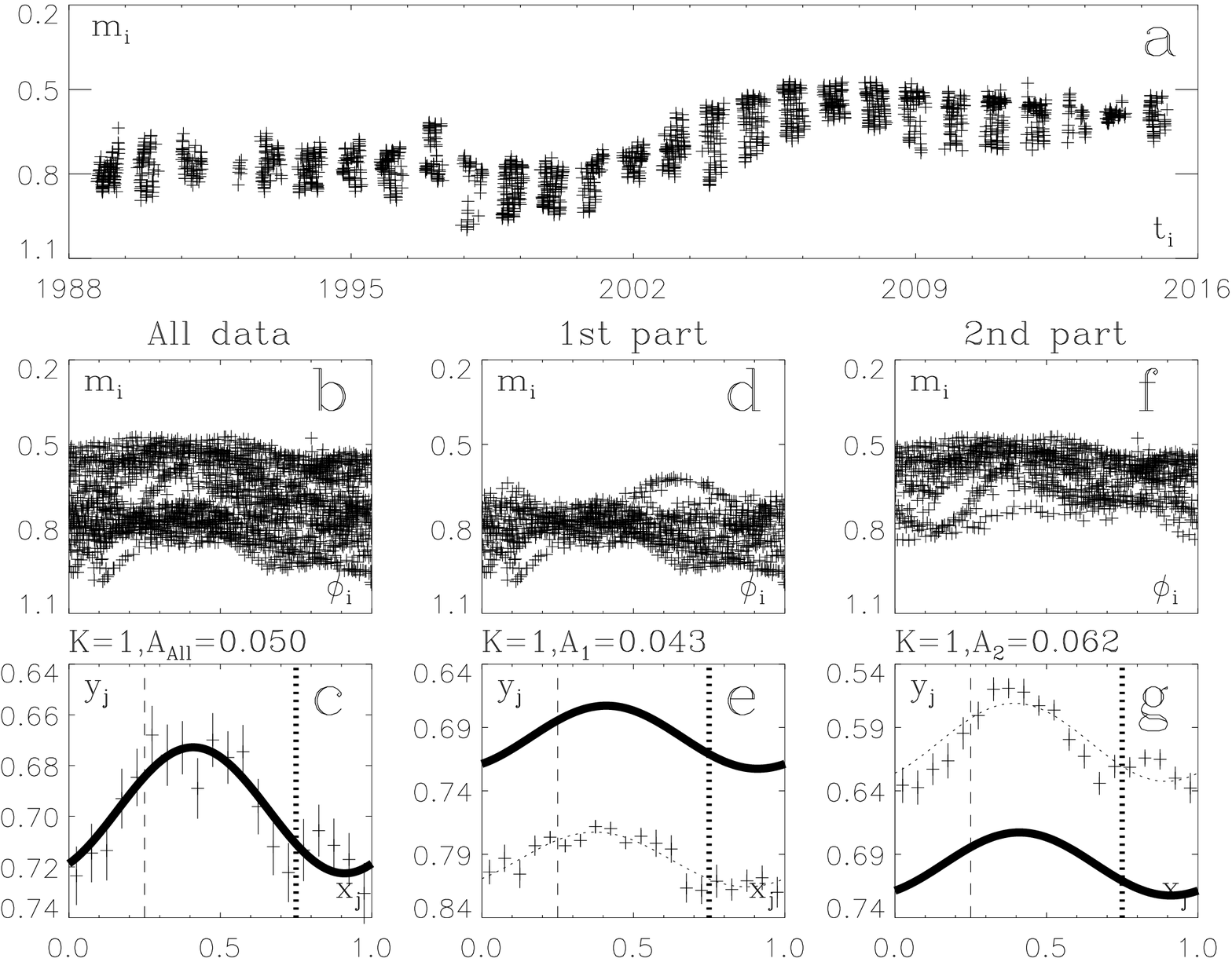}}
\caption{V1149~Ori, otherwise as in Fig. \ref{dmuma}.
\label{v1149ori}}
\end{figure}

\subsection{MLC of V1149 Ori}

\object{V1149~Ori} has a sinusoidal MLC with an amplitude of $A_{\mathrm{All}}=0.^{\mathrm{m}}050$
(Fig. \ref{v1149ori}c). 
MLC amplitude, shape and minimum are stable 
(Figs. \ref{v1149ori}ceg: $\Delta A= 0.^{\mathrm{m}}012$), 
although MLC mean changes are large (Figs. \ref{v1149ori}df).
Epochs of Ac and Aa occur about $\Delta \phi_{\mathrm{orb}}=0.15$ before 
MLC maximum and minimum 
\citep[][$P_{\mathrm{orb}}$, $t_0$, $e$, Sp-type of member A]{Hal91,Fek05}.
Note the MLC dip at about $\phi_{\mathrm{orb}}=0.75$ in Figs. \ref{v1149ori}ceg, which will
be discussed later in Sect. \ref{argumentfour}.

\begin{figure} 
\resizebox{\hsize}{!}{\includegraphics{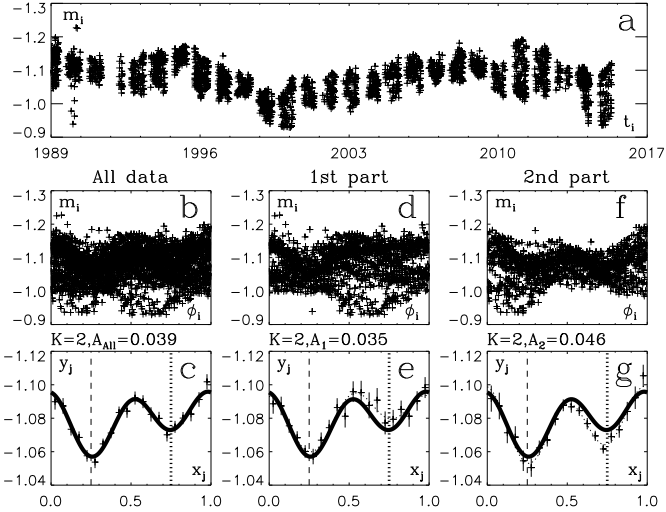}}
\caption{$\sigma$~Gem , otherwise as in Fig. \ref{dmuma}.
\label{sigmagem}}
\end{figure}

\subsection{MLC of $\sigma$ Gem}

\object{$\sigma$ Gem} has a stable double peaked MLC 
with an amplitude of $A_{\mathrm{All}}=0.^{\mathrm{m}}039$
(Fig. \ref{sigmagem}: $\Delta A=0.^{\mathrm{m}}011$).
MLC mean, amplitude, minimum and maximum do not change.
Small changes are seen only in the interval
$0.50 < \phi_{\mathrm{orb}} < 0.75$.
MLC primary and secondary minima coincide
with the Ac and Aa epochs
\citep[][$P_{\mathrm{orb}}$, $t_0$, $e$, Sp-type of member A]{Str88,Bop89,Due97}.

\begin{figure} 
\resizebox{\hsize}{!}{\includegraphics{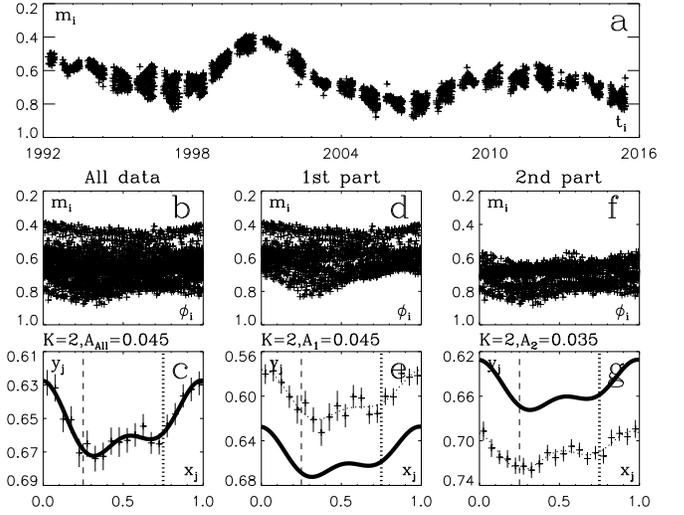}}
\caption{FG UMa , otherwise as in Fig. \ref{dmuma}.
\label{hd89546}}
\end{figure}

\subsection{MLC of FG UMa}

\object{FG UMa} ~has a stable second order MLC $(K=2)$.
MLC amplitude is about constant ($\Delta A = 0.^{\mathrm{m}}010$).
The double peaked shape and phase of this MLC 
is the same in Figs. \ref{hd89546}ceg,
although the mean levels of the 1st and 2nd
part of the $m_i$ data are different in Figs. \ref{hd89546}df.
The activity level changes alter only MLC mean,
but not MLC shape, minimum, maximum or amplitude.
MLC primary and secondary minima nearly coincide with Ac and Aa
\citep[][$P_{\mathrm{orb}}$, $t_0$, $e$, Sp-type of member A]{Fek02}.

\begin{figure} 
\resizebox{\hsize}{!}{\includegraphics{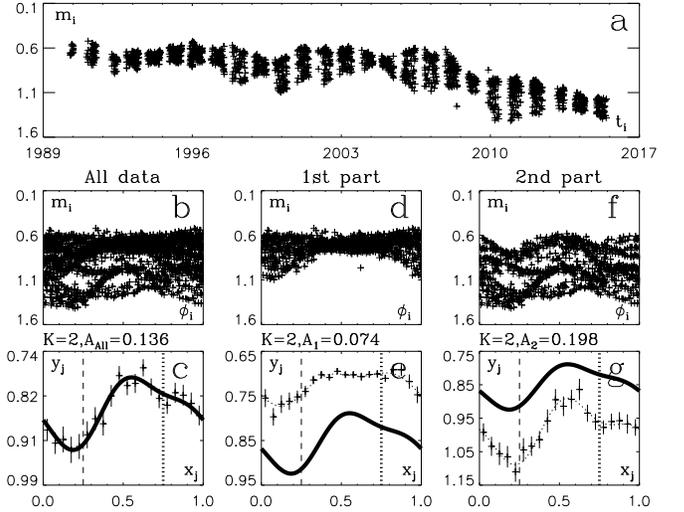}}
\caption{HU~Vir, otherwise as in Fig. \ref{dmuma}.
\label{huvir}}
\end{figure}

\subsection{MLC of HU~Vir}

\object{HU~Vir} has a high amplitude, $A_{\mathrm{All}}=0.^{\mathrm{m}}136$, second order
MLC with a nearly sinusoidal shape (Fig. \ref{huvir}c).
MLC amplitude is smaller in 1st part, $A_1=0.074$, 
and increases to an extreme value $A_2=0.198$ in the 2nd part.
The Ac epoch is close to MLC primary minimum, 
while Aa is close to a weak secondary minimum
which is more clearly visible in Fig. \ref{huvir}g
of the 2nd part
\citep[][$P_{\mathrm{orb}}$, $t_0$, $e$, Sp-type of member A]{Cut93,Fek99}.

\begin{figure} 
\resizebox{\hsize}{!}{\includegraphics{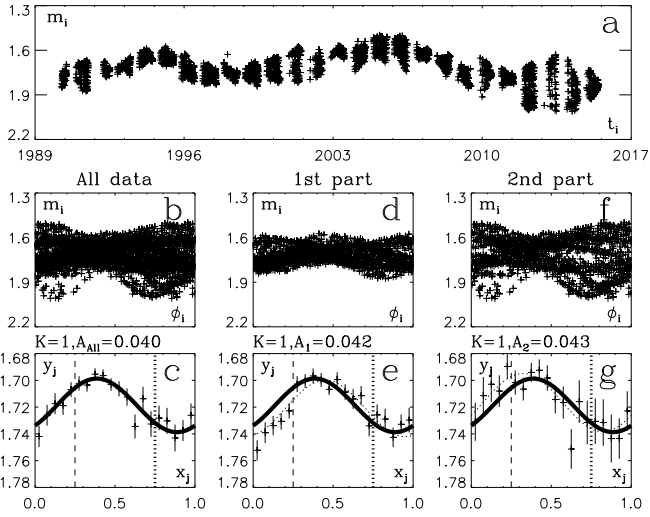}}
\caption{BM~CVn, otherwise as in Fig. \ref{dmuma}.
\label{bmcvn}}
\end{figure}

\subsection{MLC of BM~CVn}

\object{BM~CVn} has a stable sinusoidal MLC (Figs. \ref{bmcvn}cdg).
Small MLC changes occur in the  $-0.10 < \phi_{\mathrm{orb}} < 0.40$ interval,
but none in the  $0.40 < \phi_{\mathrm{orb}} < 0.90$ interval.
Aa and Ac epochs are about $\Delta \phi_{\mathrm{orb}}=0.15$ before
MLC minimum and maximum
\citep[][$P_{\mathrm{orb}}$, $t_0$, $e$, Sp-type of member A]{Gri88,Koe02}.

\begin{figure} 
\resizebox{\hsize}{!}{\includegraphics{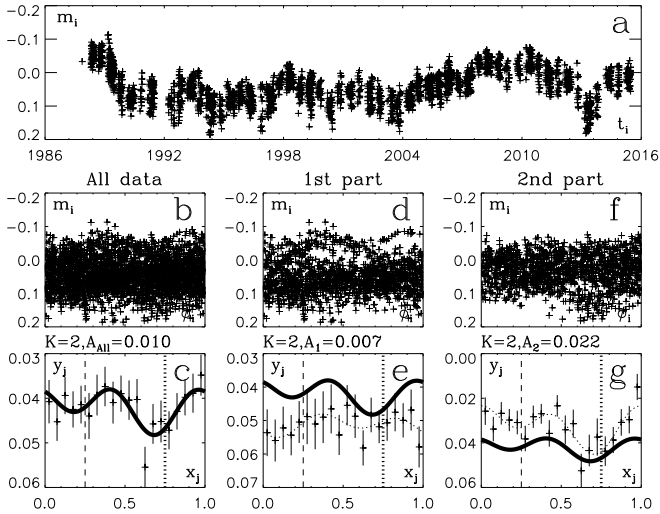}}
\caption{V478~Lyr, otherwise as in Fig. \ref{dmuma}.
\label{v478lyr}}
\end{figure}

\subsection{MLC of V478~Lyr}

MLC of \object{V478~Lyr} is double peaked and has a very low
amplitude of $A_{\mathrm{All}}=0.^{\mathrm{m}}010$. MLC shape and amplitude
do not change a lot (Figs. \ref{v478lyr}ceg: $\Delta A=0.^{\mathrm{m}}015$),
although the activity levels do  (Figs. \ref{v478lyr}df).
MLC primary minimum at $\phi_{\mathrm{orb}}=0.10$ in the 1st part shifts to
$\phi_{\mathrm{orb}}=0.65$ in the 2nd part.
The Ac and Aa epochs occur about $\Delta \phi_{\mathrm{orb}}=0.1$ after
MLC secondary and primary minima
\citep[][$P_{\mathrm{orb}}$, $t_0$, $e$, Sp-type of member A]{Gri88}.

\begin{figure} 
\resizebox{\hsize}{!}{\includegraphics{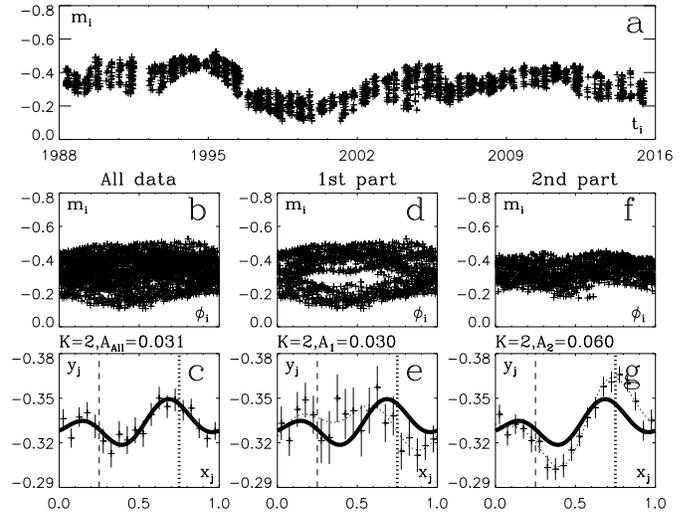}}
\caption{V1762~Cyg , otherwise as in Fig. \ref{dmuma}.
\label{v1762cyg}}
\end{figure}

\subsection{MLC of V1762~Cyg}

\object{V1762~Cyg} has a double peaked MLC with
$A_{\mathrm{All}}=0.^{\mathrm{m}}031$ (Fig. \ref{v1762cyg}c).
MLC mean is stable but MLC amplitude changes 
occur ($\Delta A=0.^{\mathrm{m}}030$).
The MLC primary minimum at $\phi_{\mathrm{orb}} \approx 0.4$ in the 1st part
shifts to $\phi_{\mathrm{orb}}=0.9$ in the 2nd part
(Figs. \ref{v1762cyg}ceg).
MLC is stable only in the short interval $0.00<\phi_{\mathrm{orb}}<0.15$.
The Ac epoch occurs about $\Delta \phi_{\mathrm{orb}}=0.15$ before
MLC primary minimum, while Aa remains close to MLC maximum 
\citep[][$P_{\mathrm{orb}}$, $t_0$, $e$, SP-type of A]{Ost98}.
Note the switch of MLC minimum and maximum in
Figs. \ref{v1762cyg}ef between the 1st and 2nd part data.
This switch will be discussed in
Sect. \ref{argumentfive}.

\begin{figure} 
\resizebox{\hsize}{!}{\includegraphics{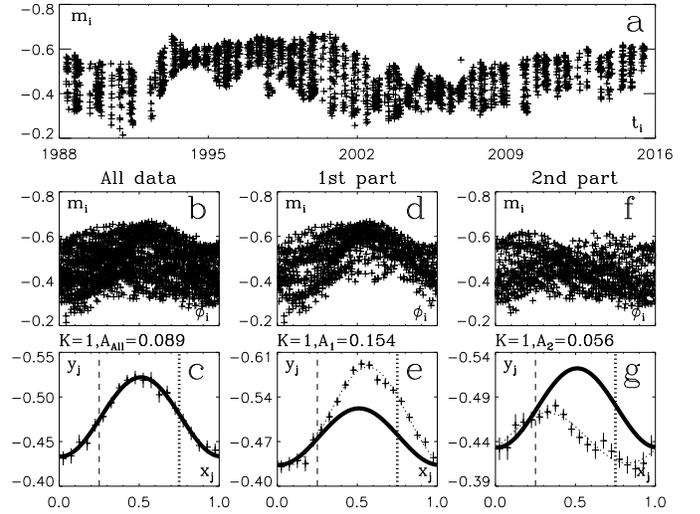}}
\caption{HK~Lac , otherwise as in Fig. \ref{dmuma}.
\label{hklac}}
\end{figure}

\subsection{MLC of HK~Lac}

\object{HK~Lac} has a high amplitude, $A_{\mathrm{All}}=0.^{\mathrm{m}}089$, sinusoidal MLC 
which remains stable at $0.00 \le \phi_{\mathrm{orb}} \le 0.30$
(Fig. \ref{hklac}cef). 
Its amplitude increases to $A_1=0.^{\mathrm{m}}154$ during the 1st part of data
and decreases to $A_2=0.^{\mathrm{m}}056$ during the 2nd part.
MLC minimum and shape remain nearly unchanged.
Epochs Ac and Aa coincide with MLC maxima and minima in Fig. \ref{hklac}g
during the 2nd part of data 
\citep[][$P_{\mathrm{orb}}$, $t_0$, $e$, Sp-type of member A]{Koe02,Gor71,Oze99,Car05}.

\begin{figure} 
\resizebox{\hsize}{!}{\includegraphics{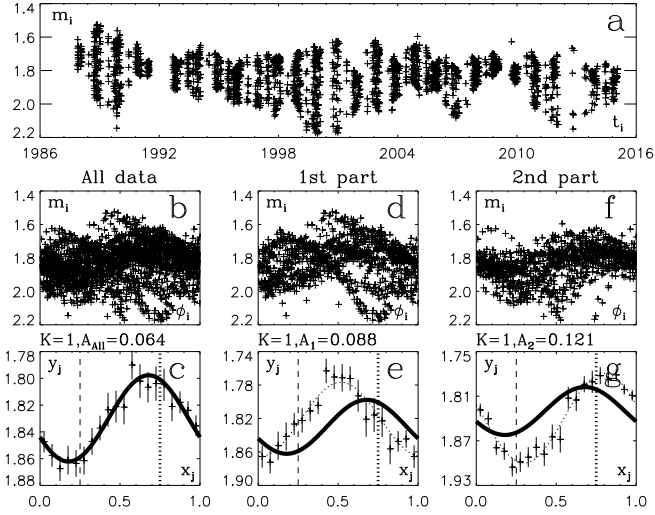}}
\caption{II~Peg, otherwise as in Fig. \ref{dmuma}.
\label{iipeg}}
\end{figure}

\subsection{MLC of II~Peg}

\object{II~Peg} ~has a sinusoidal MLC with a relatively large
amplitude, $A_{\mathrm{All}}=^{\mathrm{m}}0.064$ (Fig. \ref{iipeg}c). 
The amplitude is $A_1=0.^{\mathrm{m}}088$ in the 1st part of data,
and then increases to  $A_2=0.^{\mathrm{m}}121$ in the 2nd part.
MLC minimum and maximum phases are close to Ac and Aa in Fig. \ref{iipeg}c,
but they shift $\Delta \phi_{\mathrm{orb}}=-0.2$
backwards in the 1st part, 
and then return back to phases $\phi_{\mathrm{orb}}=0.3$ and 0.8 in the 2nd part
\citep[][$P_{\mathrm{orb}}$, $t_0$, $e$, Sp-type of member A]{Ber98}.

\section{CPS analysis results}

Next, we apply the Continuous Period Search 
\citep[][hereafter CPS]{Leh11}
to the differential $V$ magnitudes of our 14 CABS.
This gives us the following light curve parameters

\begin{figure} 
\resizebox{\hsize}{!}{\includegraphics{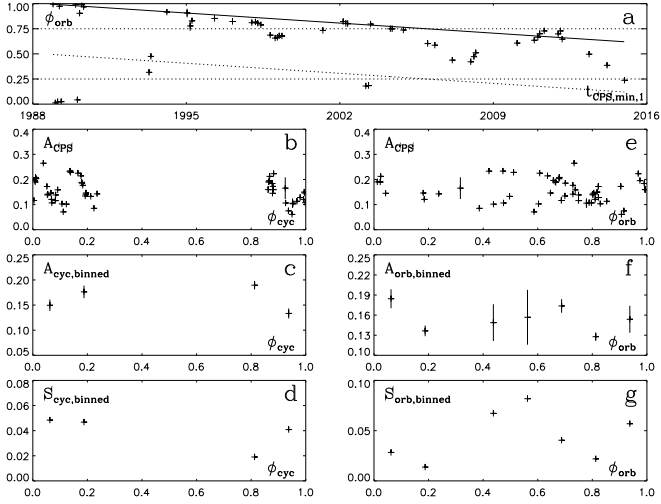}}
\caption{DM~UMa. \label{dmuma_long}
({\bf a}) Orbital phases $\phi_i$ for primary minima 
$t_{\mathrm{CPS,min,1}}$,
Orbital phases $\phi_{\mathrm{orb}}=0.25$ and 0.75 (dotted lines),
Active longitude phases $\phi_{\mathrm{orb}}=0$ (continuous line)
and 0.50 (dashed line).
({\bf b}) All amplitudes $A_{\mathrm{CPS}}$ versus $\phi_{\mathrm{cyc}}$,
({\bf c}) Binned amplitudes $A_{\mathrm{CPS,binned}}$  versus $\phi_{\mathrm{cyc}}$,
({\bf d}) Binned standard deviations $A_{\mathrm{CPS,binned}}$
 versus $\phi_{\mathrm{cyc}}$,
({\bf e}) All amplitudes $A_{\mathrm{CPS}}$ versus $\phi_{\mathrm{orb}}$,
({\bf f}) Binned amplitudes $A_{\mathrm{CPS,binned}}$  versus $\phi_{\mathrm{orb}}$,
({\bf g}) Binned standard deviations $A_{\mathrm{CPS,binned}}$
versus $\phi_{\mathrm{orb}}$}
\end{figure}

\begin{figure} 
\resizebox{\hsize}{!}{\includegraphics{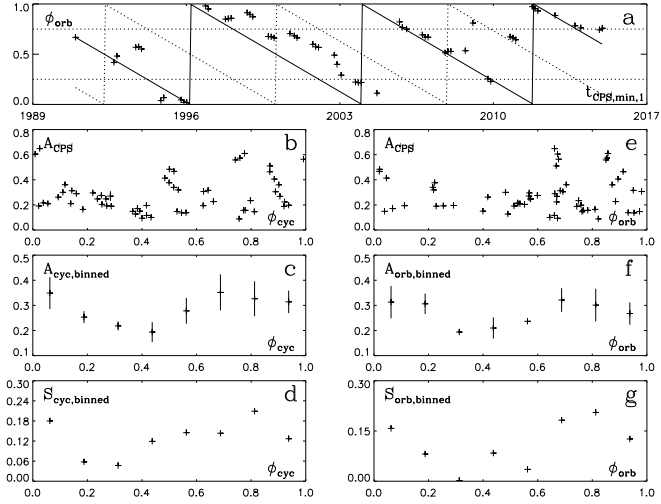}}
\caption{XX~Tri,
otherwise as in Fig. \ref{dmuma_long} 
\label{xxtri_long}}
\end{figure}

\begin{itemize}
\item[] $M_{\mathrm{CPS}}(\tau_{\mathrm{CPS}})=$ 
Mean
\item[] $A_{\mathrm{CPS}}(\tau_{\mathrm{CPS}})=$ 
Peak to peak amplitude
\item[] $P_{\mathrm{CPS}}(\tau_{\mathrm{CPS}})=$ 
Period
\item[] $t_{\mathrm{CPS,min,1}}(\tau_{\mathrm{CPS}})=$ 
Primary minimum epoch
\item[] $t_{\mathrm{CPS,tmin,2}}(\tau_{\mathrm{CPS}})=$ 
Secondary minimum epoch
\end{itemize}
\noindent
where $\tau_{\mathrm{CPS}}$ is the mean of all
observing times $t_i$ of $y_i$ of the modelled dataset.
\citet{Leh11} formulated this method.
It has been applied to the photometry numerous 
stars \citep[e.g.][]{Hac11,Kaj14,Leh16} and
it is therefore not described here in greater detail.
We use only the results for independent and reliable datasets.
``Independent'' means that the modelled datasets 
do not overlap, i.e. they have no common $y_i$ values.
The meaning of ``reliable'' is that all model parameters,
e.g. the residuals of the model,
have a Gaussian distribution.
The interesting parameters for the current study are only
$A_{\mathrm{CPS}}$ and $t_{\mathrm{CPS,min,1}}$,
because Fig. \ref{dmuma}--\ref{iipeg_long} are in particularly
connected to the changes of these two parameters.
In other words, the periods $P_{\mathrm{orb}}$ in Fig. \ref{dmuma}-\ref{iipeg} are constant,
while Figs. \ref{dmuma_long} and \ref{iipeg_long} show only the changes
of  $A_{\mathrm{CPS}}$ and $t_{\mathrm{CPS,min,1}}$.
The number of estimates obtained for these parameters are
given Table \ref{resultsthree}.
Very few reliable estimates are obtained for \object{V478~Lyr} $(n=6)$,
because its photometric rotation period is so close
to 2.1 days. It is difficult to get an adequate
phase coverage within 30 days before its light curve changes.
Therefore, the analysis of this CABS stops here.

\tablepage
\begin{table}
\begin{center}
\caption{\label{resultsthree} Active longitude.}
\addtolength{\tabcolsep}{-0.08cm}
\begin{tabular}{lllllrr}
\tableline \tableline
Variable & $n$ & $P_{\mathrm{act}}$ & $Q_{\mathrm{K}}$ & $t_{\mathrm{cyc,0}}$ &
$P_{\mathrm{cyc}}$ & $P_{\mathrm{cyc}}$ \\
         &     & $[{\mathrm{d}}]$ &               & $[{\mathrm{HJD}}]$ &
$[{\mathrm{d}}]$ & $[{\mathrm{y}}]$ \\
\tableline
      DM~UMa &   54 &    $7.4898\pm0.0008$ &   $3{\mathrm{x}}10^{-11}$ &  47499.990 &  25506 &   69.8 \\
      XX~Tri &   59 &       $23.77\pm0.01$ &   $2{\mathrm{x}}10^{-11}$ &  48237.800 &   2860 &    7.8 \\
      EL~Eri &   19 &       $47.69\pm0.02$ &                  $0.0006$ &  49236.206 &   4017 &   11.0 \\
    V711~Tau &   70 &    $2.8924\pm0.0002$ &                   $0.003$ &  47171.961 &    126 &    0.3 \\
      EI~Eri &   24 &    $1.9545\pm0.0008$ &                   $0.008$ &  47453.341 &    527 &    1.4 \\
   V1149~Ori &   26 &       $53.14\pm0.06$ &                    $0.02$ &  47467.507 &   6550 &   17.9 \\
$\sigma$~Gem &   97 &     $19.497\pm0.005$ &    $2{\mathrm{x}}10^{-8}$ &  47166.728 &   3557 &    9.7 \\
      FG~Uma &   91 &       $21.12\pm0.01$ &    $2{\mathrm{x}}10^{-6}$ &  48706.889 &   1883 &    5.2 \\
      HU~Vir &   74 &     $10.419\pm0.001$ &    $1{\mathrm{x}}10^{-8}$ &  48355.411 &   3438 &    9.4 \\
      BM~CVn &   93 &     $20.513\pm0.006$ &    $1{\mathrm{x}}10^{-6}$ &  47989.696 &   3771 &   10.3 \\
   V1762~Cyg &   64 &       $28.17\pm0.02$ &                  $0.0002$ &  47284.236 &   1919 &    5.3 \\
      HK~Lac &   67 &       $24.40\pm0.01$ &    $4{\mathrm{x}}10^{-9}$ &  47299.669 &  20988 &   57.5 \\
      II~Peg &   77 &    $6.7119\pm0.0007$ &    $8{\mathrm{x}}10^{-8}$ &  47122.096 &   3667 &   10.0 \\
\end{tabular}
\addtolength{\tabcolsep}{+0.08cm}
\tablecomments{Variable (Col 1), 
Number of $t_{\mathrm{CPS,min,1}}$ estimates (Col 2: $n$),
Active longitude period and its critical level
(Cols 2 and 3: $P_{\mathrm{act}}$, $Q_{\mathrm{K}}$),
Lowest $t_{\mathrm{CPS,min,1}}$ value (Col 4: $[t_{\mathrm{cyc,0}}])$),
Lap cycle period of Eq. \ref{cycle} (Cols 5 and 6:
$[P_{\mathrm{cyc}}]$)}

\end{center}
\end{table}

We apply the non--weighted Kuiper test formulated in \citet{Jet96A}
to the primary minima $t_{\mathrm{CPS,min,1}}$ of the remaining thirteen CABS.
The tested period interval is $\pm 15$ \% at both sides 
of $P_{\mathrm{orb}}$.
The results for the active longitude periods $(P_{\mathrm{act}})$
and their critical levels $(Q_{\mathrm{K}})$ \citep[][their Eq. 24]{Jet96A}
are given in Table \ref{resultsthree}.
The critical $(Q_{\mathrm{K}})$ levels are significant,
i.e. active longitudes definitely represent
a real dominant phenomenon
in these CABS. 
The active longitudes of
\object{V1149~Ori} have the lowest significance $(Q_{\mathrm{K}}=0.02)$.

We define a lap cycle period 
\begin{eqnarray}
P_{\mathrm{cyc}}=  |[P_{\mathrm{orb}}^{-1}-P_{\mathrm{act}}^{-1}]^{-1 }|,
\label{cycle}
\end{eqnarray}
which is the time interval when the difference in completed rounds 
is one round more or less for the periods $P_{\mathrm{orb}}$ and $P_{\mathrm{act}}$.
We use the absolute value of $P_{\mathrm{cyg}}$, because  
$P_{\mathrm{orb}}>P_{\mathrm{act}}$ gives a negative 
$P_{\mathrm{cyc}}$ value.
The active longitudes in these CABS can rotate
faster or slower than orbital motion
(Table \ref{resultsthree}, or compare
Figs. \ref{sigmagem_long} and \ref{huvir_long}).
The phases are
\begin{eqnarray} 
\phi_{\mathrm{act}}={\mathrm{FRAC}}[(t-t_{\mathrm{cyc,0}})/P_{\mathrm{act}}].
\end{eqnarray}
where $t_{\mathrm{cyc,0}}$ is
the first $t_{\mathrm{CPS,min,1}}$ value of 
each CABS given in Table \ref{resultsthree}.
The lap cycle phases are
\begin{eqnarray}
\phi_{\mathrm{cyc}}={\mathrm{FRAC}}
[(t-t_{\mathrm{cyc,0}})/P_{\mathrm{cyc}}].
\end{eqnarray}
We compute the binned $A_{\mathrm{cyc,binned}}(\phi_{\mathrm{cyc}})$ 
values of all $A_{\mathrm{CPS}}$ amplitudes with respect to this phase.
The standard deviations of these binned amplitudes within
each bin, $S_{\mathrm{cyc,binned}}(\phi_{\mathrm{cyc}})$, are also computed.
These standard deviations measure the scatter of
$A_{\mathrm{CPS}}$ values within each bin.
We use only eight bins, because the total number of 
$A_{\mathrm{CPS}}$ estimates is low, typically about fifty values.
We also compute the binned $A_{\mathrm{orb,binned}}(\phi_{\mathrm{orb}})$ 
values with respect to the orbital phase $\phi_{\mathrm{orb}}$,
as well as the standard deviations
$S_{\mathrm{orb,binned}}(\phi_{\mathrm{orb}})$. 
The scatter can not be large 
for amplitudes very close to zero,
and this may introduce some bias in the interpretation
of $S_{\mathrm{cyc,binned}}$ and $S_{\mathrm{orb,binned}}$ changes.
The results for our thirteen CABS are shown in
Figs. \ref{dmuma_long} -- \ref{iipeg_long}.

\subsection{Active longitudes and amplitudes  of  DM~UMa} 
\label{dmumaactive}

The lap cycle of \object{DM~UMa} ~is surprisingly long, 
$P_{\mathrm{cyc}}=69.8$ years!
The regular migration of the $t_{\mathrm{CPS,min,1}}$ phases
becomes irregular when the active longitude crosses
the orbital period phase $\phi_{\mathrm{orb}}=0.75$ 
(Fig. \ref{dmuma_long}a: crossing continuous and dotted lines).
The gap with no data in Fig. \ref{dmuma_long}b
awaits for the missing future observations.
The connection $A_{\mathrm{CPS}}$ to $\phi_{\mathrm{orb}}$ is
not clear (Figs. \ref{dmuma_long}b).
We get only four binned $A_{\mathrm{cyc,binned}}(\phi_{\mathrm{cyc}})$
and $S_{\mathrm{cyc,A}}(\phi_{\mathrm{cyc}})$ values
(Figs. \ref{dmuma_long}cd).
The seven $A_{\mathrm{orb,binned}}$ values
(Fig. \ref{dmuma_long}f) show a regular connection to 
$\phi_{\mathrm{orb}}$ which is confirmed by the 
$S_{\mathrm{orb,binned}}$ changes
(Figs. \ref{dmuma_long}g).

\subsection{Active longitudes and amplitudes  of XX~Tri} 
\label{xxtriactive}

The lap cycle of \object{XX~Tri} is $P_{\mathrm{cyc}}=7.8$ years.
It is clearly not the cycle of the
mean brightness (Fig. \ref{xxtri}a).
The active longitude is very stable (Fig. \ref{xxtri_long}a).
There are some migration irregularities,  
especially when the active longitude migrates across orbital phases 
$\phi_{\mathrm{orb}}=0.25$ and 0.75.
The changes of $A_{\mathrm{CPS}}(\phi_{\mathrm{cyc}})$ are very regular,
as well as those of $A_{\mathrm{cyc,binned}}(\phi_{\mathrm{cyc}})$
and $S_{\mathrm{cyc,A}}(\phi_{\mathrm{cyc}})$
(Figs. \ref{xxtri_long}bcd).
The $A_{\mathrm{CPS}}(\phi_{\mathrm{orb}})$,
$A_{\mathrm{orb,binned}}(\phi_{\mathrm{orb}})$ 
and $S_{\mathrm{orb,A}}(\phi_{\mathrm{orb}})$
changes are also regular (Figs. \ref{xxtri_long}feg).

\begin{figure} 
\resizebox{\hsize}{!}{\includegraphics{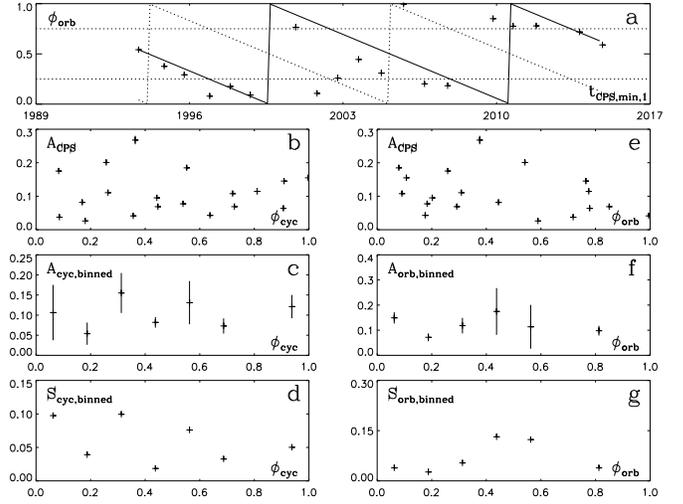}}
\caption{EL~Eri, 
otherwise as in Fig. \ref{dmuma_long} 
\label{eleri_long}}
\end{figure}

\subsection{Active longitudes and amplitudes  of  EL~Eri} 
\label{eleriactive}

Very few $t_{\mathrm{CPS,min,1}}$ and $A_{\mathrm{CPS,min,1}}$
estimates of \object{EL~Eri} are available $(n=19)$.
The active longitude migration in the $t_{\mathrm{CPS,min,1}}$ phases 
is stable Fig. (\ref{eleri_long}a).
Too few  $A_{\mathrm{CPS,min,1}}$ are available to confirm
the $P_{\mathrm{cyc}}=10.9$ years lap cycle (Figs. \ref{eleri_long}b--d). 
However, some regularity is present in $A_{\mathrm{orb,binned}}$
and $S_{\mathrm{orb,binned}}$ changes (Figs. \ref{eleri_long}f--g).

\begin{figure} 
\resizebox{\hsize}{!}{\includegraphics{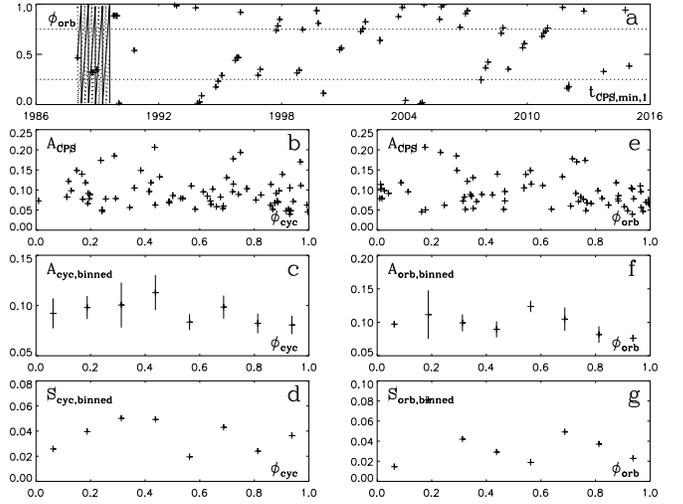}}
\caption{V711~Tau, 
otherwise as in Fig. \ref{dmuma_long} 
\label{v711tau_long}}
\end{figure}

\subsection{Active longitudes and amplitudes  of  V711~Tau} 
\label{v711tauactive}

The $P_{\mathrm{cyc}}$ lap cycle of \object{V711~Tau}
is very short, only 126 days.
The predicted migration is so fast that the
dashed and continuous lines used for
illustrating it would totally cover
Fig. \ref{v711tau_long}a,
and we therefore show this migration
only for the first five lap cycles.
The $A_{\mathrm{cyc,binned}}$ and
$C_{\mathrm{cyc,binned}}$ changes 
follow the $\phi_{\mathrm{cyc}}$ phases.
The connection of $A_{\mathrm{orb,binned}}$ and
$S_{\mathrm{orb,binned}}$ to the $P_{\mathrm{orb}}$ 
phases is also excellent.

\begin{figure} 
\resizebox{\hsize}{!}{\includegraphics{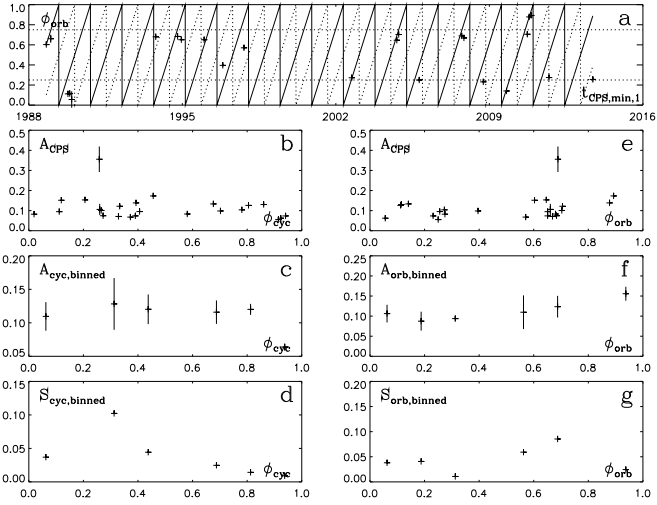}}
\caption{EI~Eri, 
otherwise as in Fig. \ref{dmuma_long} 
\label{eieri_long}}
\end{figure}

\subsection{Active longitudes and amplitudes  of  EI~Eri} 
\label{eieriactive}

The number of $t_{\mathrm{CPS,min,1}}$ and $A_{\mathrm{CPS,min,1}}$
estimates available for \object{EI~Eri} is only $n=24$.
The migration lines in Fig. \ref{eieri_long}a are very steep, 
because the $P_{\mathrm{cyc}}$ lap cycle is only 527 days long.
Most of the phases of $t_{\mathrm{CPS,min,1}}$ are 
close to $\phi_{\mathrm{orb}}=0.25$ and 0.75.
The five $A_{\mathrm{cyc,binned}}$, $S_{\mathrm{cyc,binned}}$, 
$A_{\mathrm{orb,binned}}$ or $S_{\mathrm{orb,binned}}$ estimates 
can not be used to tell anything about 
the connection to $\phi_{\mathrm{cyc}}$ or $\phi_{\mathrm{orb}}$.

\begin{figure} 
\resizebox{\hsize}{!}{\includegraphics{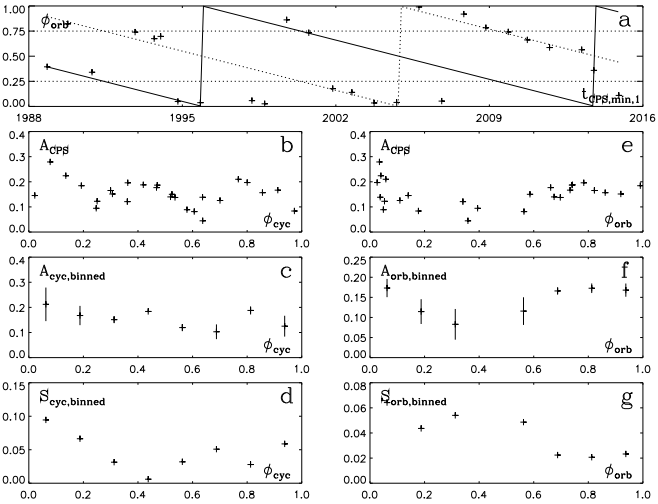}}
\caption{V1149~Ori, 
otherwise as in Fig. \ref{dmuma_long} 
\label{v1149ori_long}}
\end{figure}

\subsection{Active longitudes and amplitudes  of  V1149~Ori} 
\label{eieriactive}

The migration of the phases of
$t_{\mathrm{CPS,min,1}}$ is very stable 
for \object{V1149~Ori} (Fig. \ref{v1149ori_long}a).
Two \flip ~events occur in the years 1993 and 2000
when the dashed migration line crosses the orbital
phase $\phi_{\mathrm{orb}}=0.25$.
Even the connection of {\it individual} $A_{\mathrm{CPS}}$
estimates to $\phi_{\mathrm{cyc}}$ or $\phi_{\mathrm{orb}}$
is regular.
The $A_{\mathrm{cyc,binned}}$ and $S_{\mathrm{cyc,binned}}$
changes follow $\phi_{\mathrm{cyc}}$,
and those  of $A_{\mathrm{orb,binned}}$ and 
$S_{\mathrm{orb,binned}}$ follow $\phi_{\mathrm{orb}}$.

\begin{figure} 
\resizebox{\hsize}{!}{\includegraphics{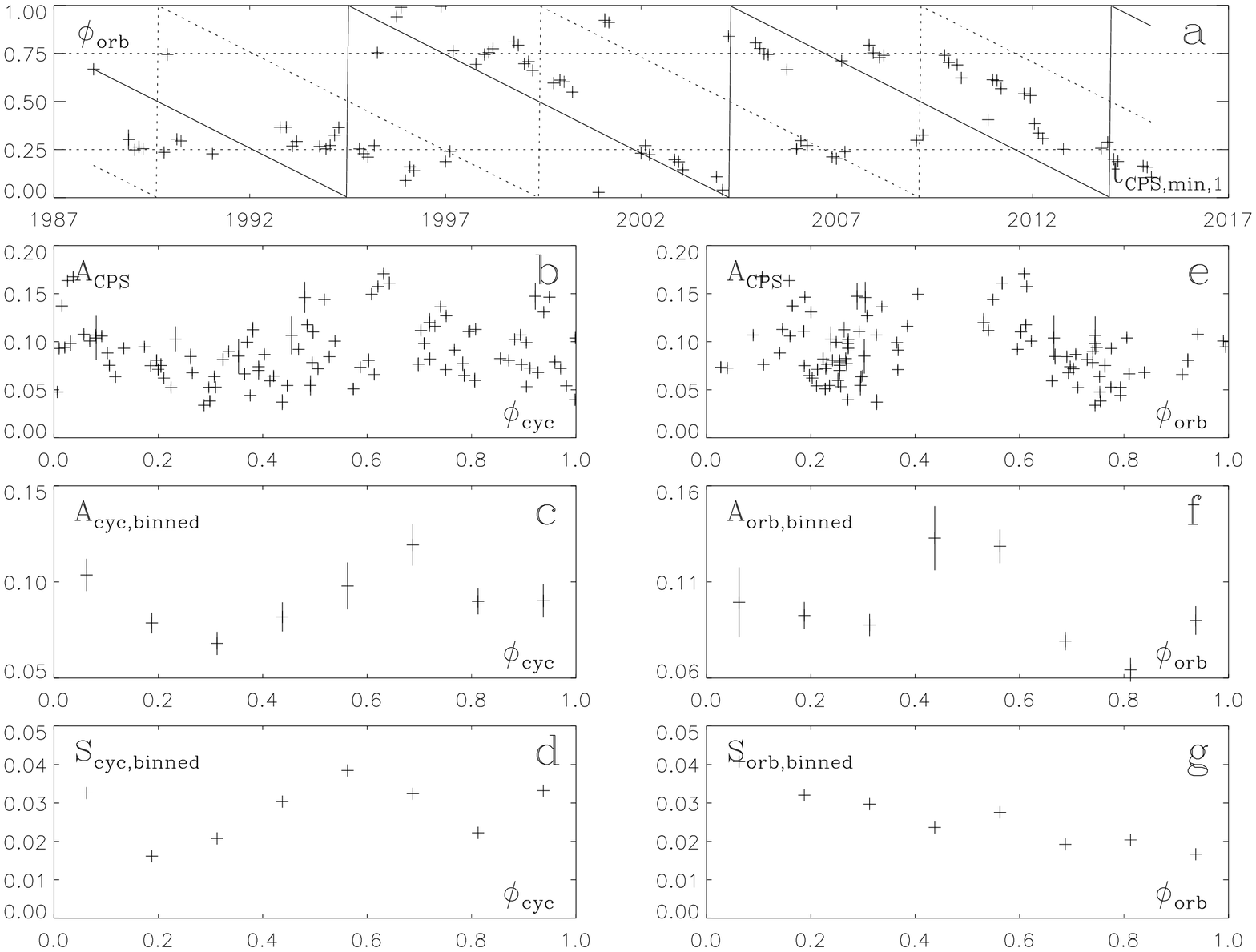}}
\caption{$\sigma$~Gem, 
otherwise as in Fig. \ref{dmuma_long} 
\label{sigmagem_long}}
\end{figure}

\subsection{Active longitudes and amplitudes  of  $\sigma$ Gem} 
\label{sigmagemsect}

How about our favourite
star, \object{$\sigma$ Gem} \citep[][``Behaves well'']{Jet96}.
Migration has remained regular (Fig. \ref{sigmagem_long}a).
Nearly all minima stayed at $\phi_{\mathrm{orb}}=0.25$ before the year 1992,
and then they began to migrate downwards. 
This migration still continues.
Four \flip ~events occur in the years 1996, 2001, 2005 and 2010. 
The changes of individual $A_{\mathrm{CPS}}$ follow
$\phi_{\mathrm{cyc}}$ (Fig. \ref{sigmagem_long}b).
Both $A_{\mathrm{cyc,binned}}$ and $S_{\mathrm{cyc,binned}}$
follow $\phi_{\mathrm{cyc}}$ beyond all ecpectations (Figs. \ref{sigmagem_long}cd).
Individual $A_{\mathrm{CPS}}$ estimates also follow $\phi_{\mathrm{orb}}$
and the largest scatter coincides with $\phi_{\mathrm{orb}}=0.25$ and 0.75
(Figs. \ref{sigmagem_long}e).
Changes of $A_{\mathrm{orb,binned}}$ and 
$S_{\mathrm{orb,binned}}$ follow $\phi_{\mathrm{orb}}$.

\begin{figure} 
\resizebox{\hsize}{!}{\includegraphics{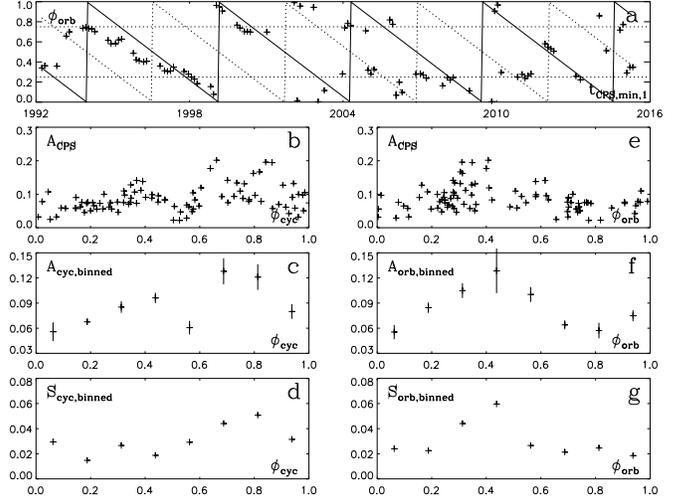}}
\caption{FG~UMa, 
otherwise as in Fig. \ref{dmuma_long} 
\label{fguma_long}}
\end{figure}

\subsection{Active longitudes and amplitudes of  FG~UMa} 
\label{fgumasect}

The migration of \object{FG~UMa} has been very regular after the
\flip ~event in the year 1994 (Fig. \ref{fguma_long}a).
The connection of 
individual $A_{\mathrm{CPS}}$ estimates to $\phi_{\mathrm{cyc}}$ is clear, 
as well as that of $A_{\mathrm{cyc,binned}}$ and $S_{\mathrm{cyc,binned}}$
(Figs. \ref{fguma_long}bcd).
The largest of $A_{\mathrm{CPS}}$
scatter coincides with $\phi_{\mathrm{orb}}=0.25$ (Fig. \ref{fguma_long}e).
The $A_{\mathrm{orb,binned}}$ and 
$S_{\mathrm{orb,binned}}$ changes are connected to
$\phi_{\mathrm{orb}}$ (Figs. \ref{fguma_long}fg).

\begin{figure} 
\resizebox{\hsize}{!}{\includegraphics{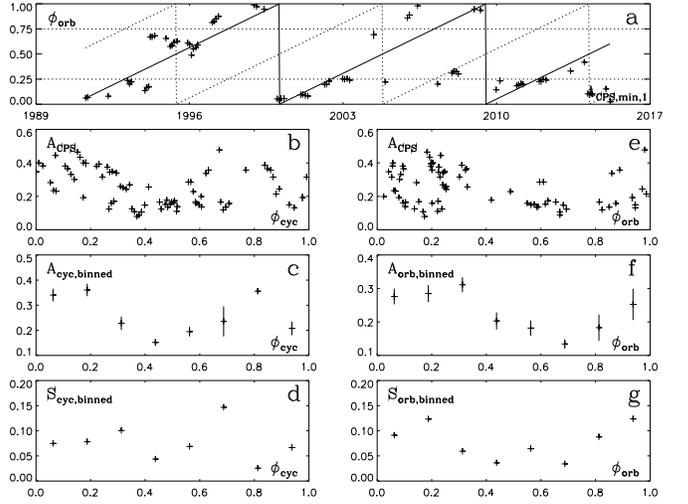}}
\caption{HU~Vir, 
otherwise as in Fig. \ref{dmuma_long} 
\label{huvir_long}}
\end{figure}

\subsection{Active longitudes and amplitudes of HU~Vir} 
\label{huvirsect}

A \flip ~occured in \object{HU~Vir} in the year 1994.
The light curve minima were at $\phi_{\mathrm{orb}}=0.75$
between the years 1994 and 1997.
A shift to $\phi_{\mathrm{orb}}=0.25$ took place in the year 1999,
and after that the $t_{\mathrm{CPS,min,1}}$ phases 
have not changed at all (Figs. \ref{huvir_long}a).  
Changes of individual $A_{\mathrm{CPS}}$ estimates,
as well as those of$A_{\mathrm{cyc,binned}}$ and $S_{\mathrm{cyc,binned}}$,
are regular (Figs. \ref{huvir_long}bcd). 
The scatter of $A_{\mathrm{CPS}}$ is largest at $\phi_{\mathrm{orb}}=0.25$ 
(Fig. \ref{huvir_long}e).
The connection of $A_{\mathrm{orb,binned}}$ and 
$S_{\mathrm{orb,binned}}$ to $\phi_{\mathrm{orb}}$ 
is clear (Figs. \ref{huvir_long}fg).

\begin{figure} 
\resizebox{\hsize}{!}{\includegraphics{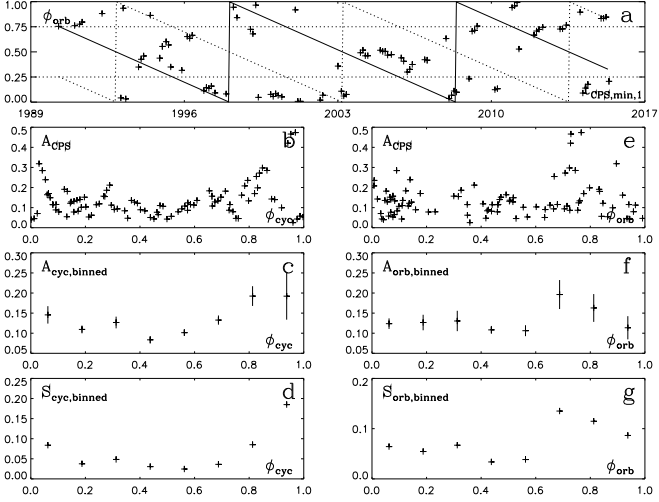}}
\caption{BM~CVn, 
otherwise as in Fig. \ref{dmuma_long} 
\label{bmcvn_long}}
\end{figure}

\subsection{Active longitudes and amplitudes of BM~CVn} 
\label{bmcvnsect}

The migration in \object{BM~CVn} is regular,
and \flip ~events occur, e.g in the years 1999 and 2009 and 2015
when the continuous line of migration crosses the dotted line
at $\phi_{\mathrm{orb}}=0.25$
(Figs. \ref{bmcvn_long}a).  
The changes of individual $A_{\mathrm{CPS}}$ estimates
display some regularity (Figs. \ref{bmcvn_long}b).
The connection of $A_{\mathrm{cyc,binned}}$ and $S_{\mathrm{cyc,binned}}$
to $\phi_{\mathrm{cyc}}$ is clear (Figs. \ref{bmcvn_long}cd).
The largest scatter of $A_{\mathrm{CPS}}$ coincides
with $\phi_{\mathrm{orb}}=0.75$ 
(Fig. \ref{bmcvn_long}e).
The $A_{\mathrm{orb,binned}}$ and 
$S_{\mathrm{orb,binned}}$ changes follow $\phi_{\mathrm{orb}}$ 
(Figs. \ref{bmcvn_long}fg).

\begin{figure} 
\resizebox{\hsize}{!}{\includegraphics{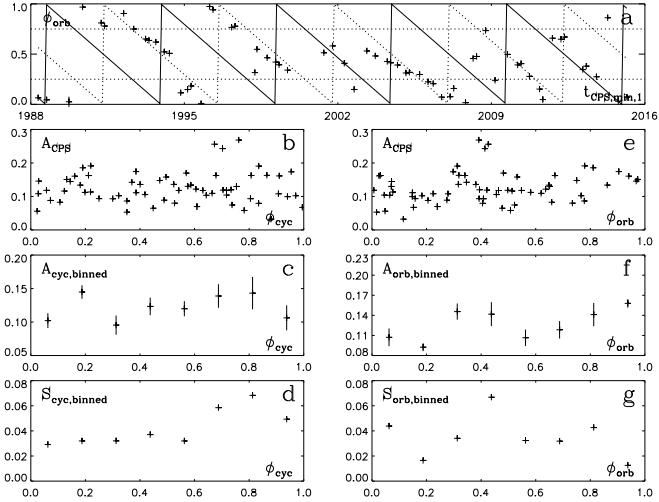}}
\caption{V1762~Cyg, 
otherwise as in Fig. \ref{dmuma_long} 
\label{v1762cyg_long}}
\end{figure}

\subsection{Active longitudes and amplitudes of V1762~Cyg} 
\label{v1762cygsect}

The clearest case of \flip ~in \object{V1762~Cyg} occurs in 2008
when the continuous line of migration crosses the orbital phase
$\phi_{\mathrm{orb}}=0.25$ (Figs. \ref{v1762cyg_long}a).  
Individual $A_{\mathrm{CPS}}$ change regularly (Figs. \ref{v1762cyg_long}b).
The $A_{\mathrm{cyc,binned}}$ and $S_{\mathrm{cyc,binned}}$ changes follow
$\phi_{\mathrm{cyc}}$ (Figs. \ref{v1762cyg_long}cd).
The scatter $A_{\mathrm{CPS}}$ is largest at $\phi_{\mathrm{orb}}=0.75$ 
(Fig. \ref{v1762cyg_long}e).
The $A_{\mathrm{orb,binned}}$ and $S_{\mathrm{orb,binned}}$ changes are also regular 
(Figs. \ref{v1762cyg_long}fg).

\begin{figure} 
\resizebox{\hsize}{!}{\includegraphics{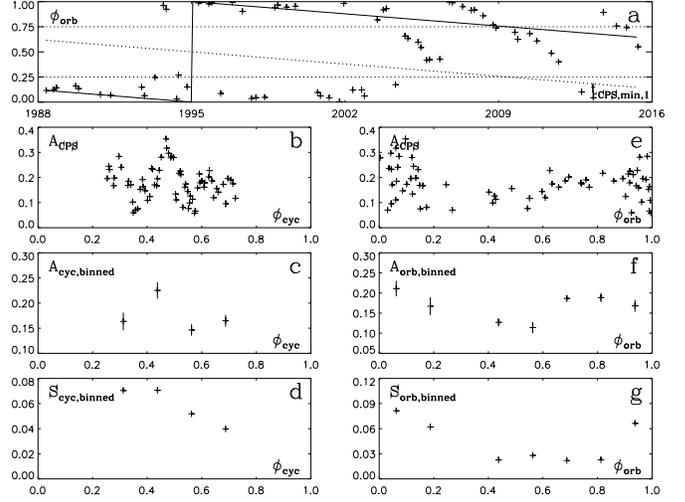}}
\caption{HK~Lac, 
otherwise as in Fig. \ref{dmuma_long} 
\label{hklac_long}}
\end{figure}

\subsection{Active longitudes and amplitudes of HK~Lac} 
\label{hklacsect}

The changes of the light curve minima of \object{HK~Lac} 
are exceptional (Figs. \ref{hklac_long}a).
The $t_{\mathrm{CPS,min,1}}$ phases remained fixed at
$\phi_{\mathrm{orb}}=0.75$ between the years 1988 and 2004.
Then linear migration began, and it has continued since then.
One \flip ~occurs in the year 2007.
The long lap cycle period of \object{HK~Lac},
$P_{\mathrm{cyc}}=57.5$ years, 
is the reason for the $\phi_{\mathrm{cyc}}$
gap with no $A_{\mathrm{CPS}}$,
$A_{\mathrm{cyc,binned}}$ and 
$S_{\mathrm{cyc,binned}}$
estimates in Figs. \ref{hklac_long}bcd.
The largest scatter of $A_{\mathrm{CPS}}$ 
occurs close to $\phi_{\mathrm{orb}}=0.75$ 
(Fig. \ref{hklac_long}e).
The changes of $A_{\mathrm{orb,binned}}$ and $S_{\mathrm{orb,binned}}$ 
are regular 
(Figs. \ref{v1762cyg_long}fg).

\begin{figure} 
\resizebox{\hsize}{!}{\includegraphics{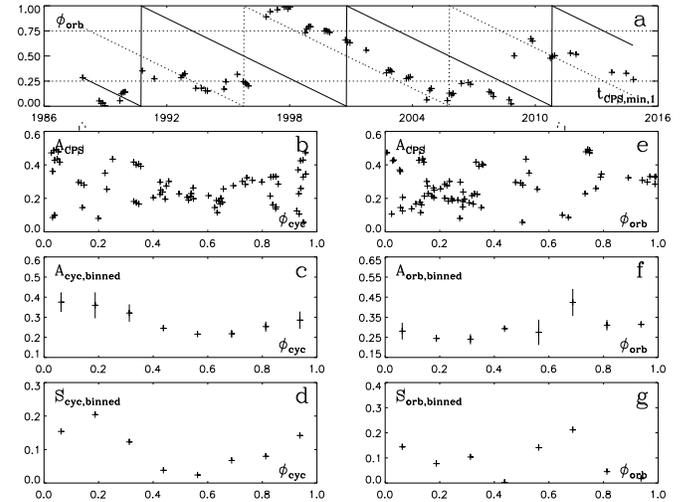}}
\caption{II~Peg, 
otherwise as in Fig. \ref{dmuma_long} 
\label{iipeg_long}}
\end{figure}

\subsection{Active longitudes and amplitudes of II~Peg} 
\label{iipegsect}

\object{II~Peg} is another famous \flip ~star \citep{Ber98A}.
Our analysis does not reveal a single clear case 
of \flip ~in this CABS (Fig. \ref{iipeg_long}a).
The $t_{\mathrm{CPS,min,1}}$ phases remained fixed at
$\phi_{\mathrm{orb}}=0.25$ between the years 1988 and 1997.
Then an extremely regular linear migration began,
and it has continued since then. 
This reveals how subjective the indentifcation of \flip ~events
can be \citep[e.g.][their Fig. 1]{Ber98A}.
The changes of individual $A_{\mathrm{CPS}}$ estimates are 
not very regular (Figs. \ref{iipeg_long}b),
while those of $A_{\mathrm{cyc,binned}}$ and $S_{\mathrm{cyc,binned}}$ 
certainly are (Figs. \ref{iipeg_long}cd).
The largest scatter of $A_{\mathrm{CPS}}$ 
coincides with $\phi_{\mathrm{orb}}=0.75$ 
(Figs. \ref{iipeg_long}eg).
The changes of $A_{\mathrm{orb,binned}}$ and $S_{\mathrm{orb,binned}}$ 
are regular 
(Figs. \ref{iipeg_long}fg).

\section{Discussion } \label{discussion}

Here, we show that
a stationary part of the light curve explains
MLC similarities and differences of our fourteen CABS
(Sect. \ref{stationary}),
while a nonstationary part of the light curve explains the  active longitudes 
(Sect. \ref{nonstationary}).

\begin{figure*}
\resizebox{\hsize}{!}{\includegraphics{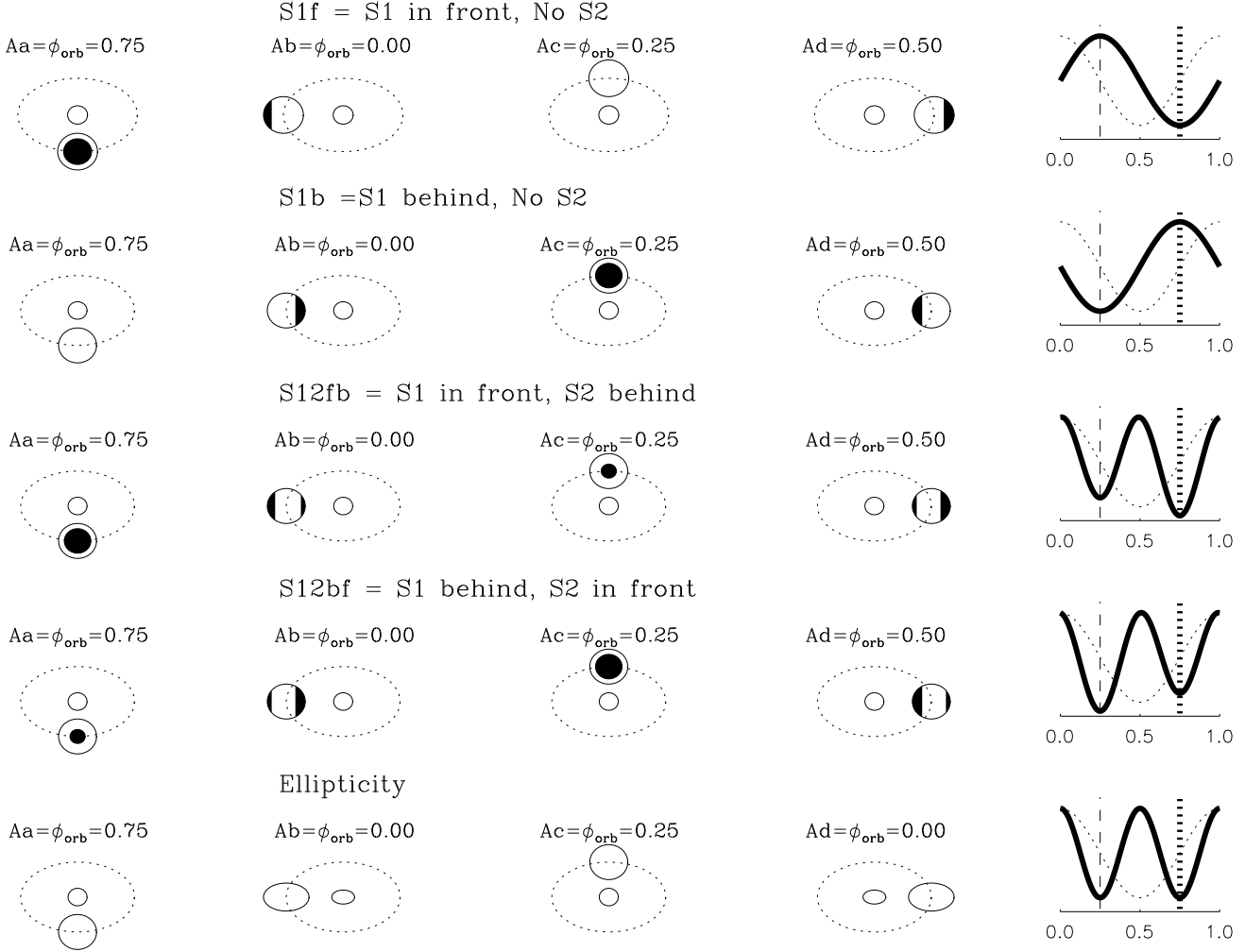}}
\caption{ \label{GFfig} ``Stationary \flip''.
The highest line: Four first sketches show the S1f mode configuration
of CABS at epochs Aa, Ab, Ac and Ad described in Sect. 
\ref{stationary}.
The fifth sketch shows the
qualitative changes of radial velocity (dashed line) and MLC (thick
continuous line) in an arbitrary scale.
Vertical lines indicate epochs Aa 
(thick dotted line) and Ac (thin dashed line).
2nd, 3rd and 4th highest lines: S1b, S12fg and S12bf mode
configurations, otherwise as in the highest line.
Lowest line: Ellipticy configurations at Aa, Ab, Ac and Ad epoch.
Radial velocity and MLC curves are as in the highest line.
\label{genflip}}
\end{figure*}

\begin{figure*}
\resizebox{\hsize}{!}{\includegraphics{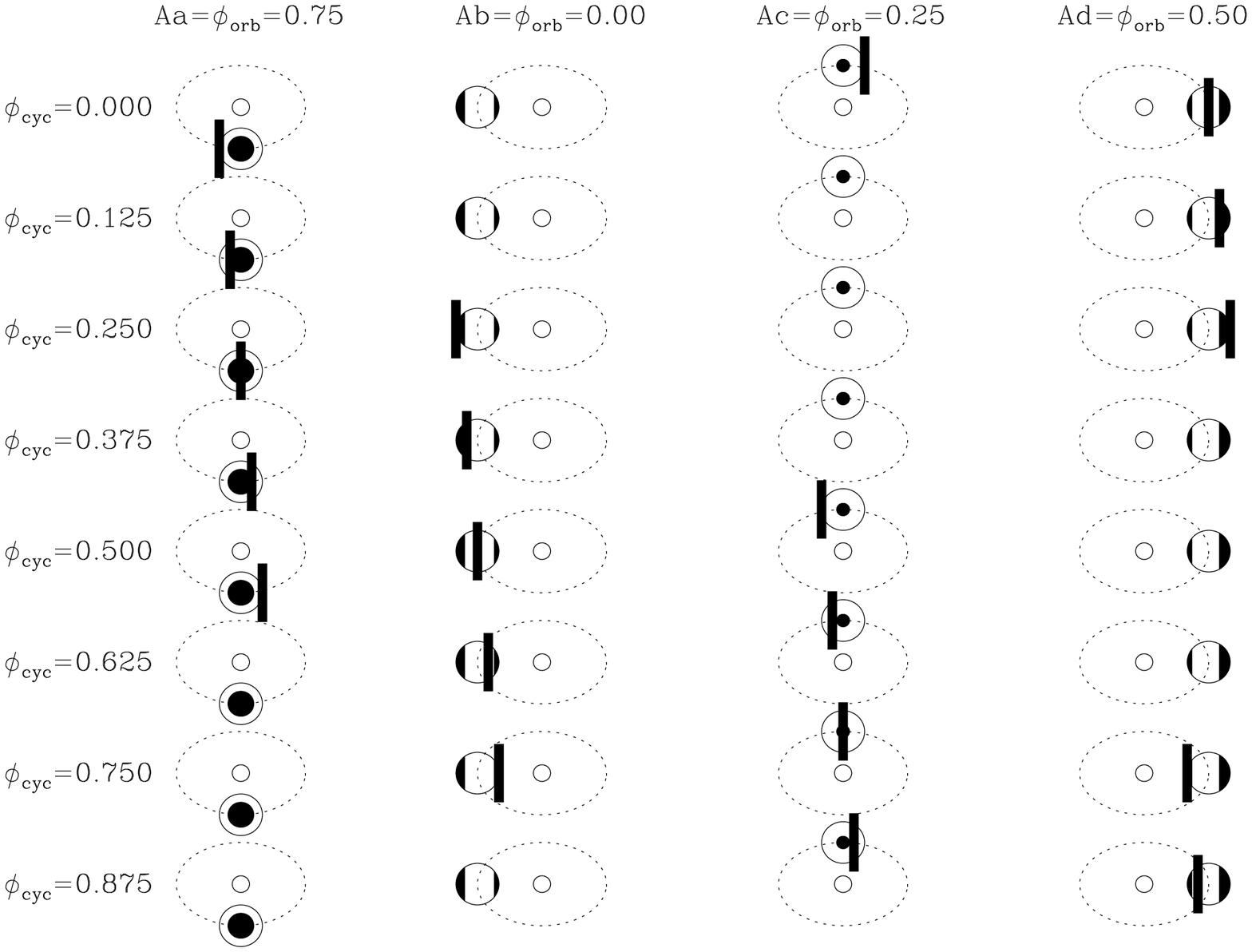}}
\caption{ \label{Myway} One complete $P_{\mathrm{cyc}}$ lap cycle.
Lap cycle phases $\phi_{\mathrm{cyc}}$ are given 
on the left of each line.
Nonstationary spot S3 (Eq. \ref{Eqnonstationary}) rotates
once around member A during $P_{\mathrm{Cyc}}$. 
Thick dark vertical line denotes the longitude of spot S3 on member A.
We mark this longitude only when spot S3 
is visible on stellar disk of member A.
Locations of stationary spots S1 and S2 (Eq. \ref{Eqstationary})
are as on the third line of Fig. \ref{genflip}.}
\end{figure*}

\subsection{Stationary part} \label{stationary}

Let us assume that the rotation of member A is synchronized
with its orbital motion around member B. In this case,
member A always turns the same side towards member B,
like the Moon always turns the same side towards the Earth.

In the 1st mode, there is only one spot S1 on A.
The brighter side of A is always turned towards B.
We observe the maximum projected area of S1 when the line connecting A and B 
is parallel to our line of sight at epoch Aa.
Our notation for this 1st mode is S1f in Fig. \ref{GFfig}
(the highest line).
MLC minimum coincides with Aa epoch in S1f.

In the 2nd mode, there is 
again only one spot S1 on A, 
but the brighter side of A is always turned away from B.
The projected area of S1 towards us is largest
when the line connecting A and B is parallel to 
our line of sight at Ac epoch.
We denote this 2nd mode as S1b in Fig. \ref{GFfig}
(the 2nd highest line).
MLC primary minimum coincides with Ac epoch in S1b mode.

There is one larger spot S1 
and one smaller spot S2 on A in the 3rd mode.
S1 is always turned away from B,
and S2 is always turned towards B. 
We observe the maximum projection of S1
when the line connecting A and B 
is parallel to our line of sight at epoch Aa.
Our notation for this 3rd mode is 
S12fb in Fig. \ref{GFfig}
(3rd highest line).
MLC primary minimum coincides with Aa epoch
and MLC secondary minimum with Ac in this S12fb mode.

In the last 4th mode,
the roles of S1 and S2 spots are reversed,
if compared to the 3rd mode,
i.e. both spots have shifted by 180 degrees.
S1 is always turned towards B and S1 always away from B.
We see the largest projected area of S1 at epoch Ac
when S2 is out of sight.
Our notation for this 4th mode is 
S12bf in Fig. \ref{GFfig}
(2nd lowest line).
MLC primary and secondary minima coincide 
with Ac and Aa in this S12bf mode.

The lowest line of Fig. \ref{GFfig} shows
how ellipticity causes two identical MLC minima at Aa and Ac.

The direction of orbital motion 
is fixed from left (Aa) to right (Ab) in Fig. \ref{GFfig}.
The results would be same, if this direction were from right to left,
because all $e$ values our CABS are zero, or close to zero
(Table \ref{resultsone}).

Comparison of MLC in Figs. \ref{dmuma} -- \ref{iipeg} 
to Fig. \ref{GFfig} reveals the modes of our fourteen CABS:
five in S1f    mode  (\object{DM~UMa}, \object{XX~Tri}, \object{V1149~Ori}, 
\object{BM~CVn} , \object{HK~Lac}),
two  in S1b    mode  (\object{EL~Eri}, \object{II~Peg}),
two  in S12fb  mode  (\object{EI~Eri}, \object{V478~Lyr}) and
five in S12bf  mode (\object{V711~Tau}, \object{$\sigma$ Gem}, \object{FG UMa}, \object{HU~Vir}, \object{V1762~Cyg}).

The evidence for this \genflip ~model is overwhelming,
as explained in the arguments below.

\begin{figure*} 
\resizebox{\hsize}{!}{\includegraphics{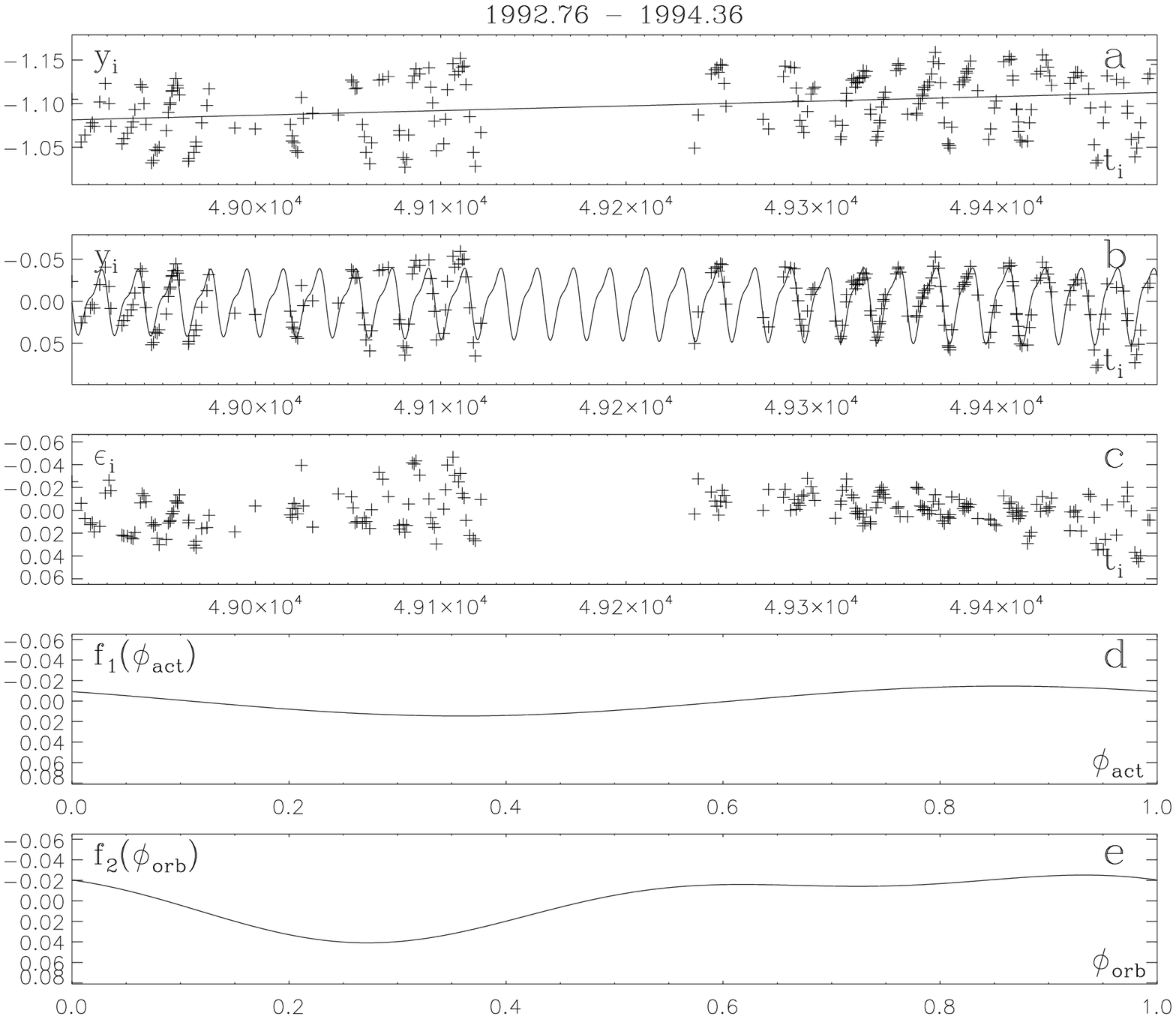}}
\caption{Arbitrary slice of $\sigma$~Gem photometry.
({\bf a}) Original photometry and a linear fit,
({\bf b}) Original photometry minus linear fit,
and model of Eq. \ref{solution} $(K_1=1, K_2=2, M=M_1=M_2=0)$,
({\bf c}) Model residuals $\epsilon_i$,
({\bf d}) Nonstationary part $f_1(\phi_{\mathrm{act}})$,
({\bf e}) Stationary part $f_2(\phi_{\mathrm{orb}})$
\label{sliceone}}
\end{figure*}

\begin{figure*} 
\resizebox{\hsize}{!}{\includegraphics{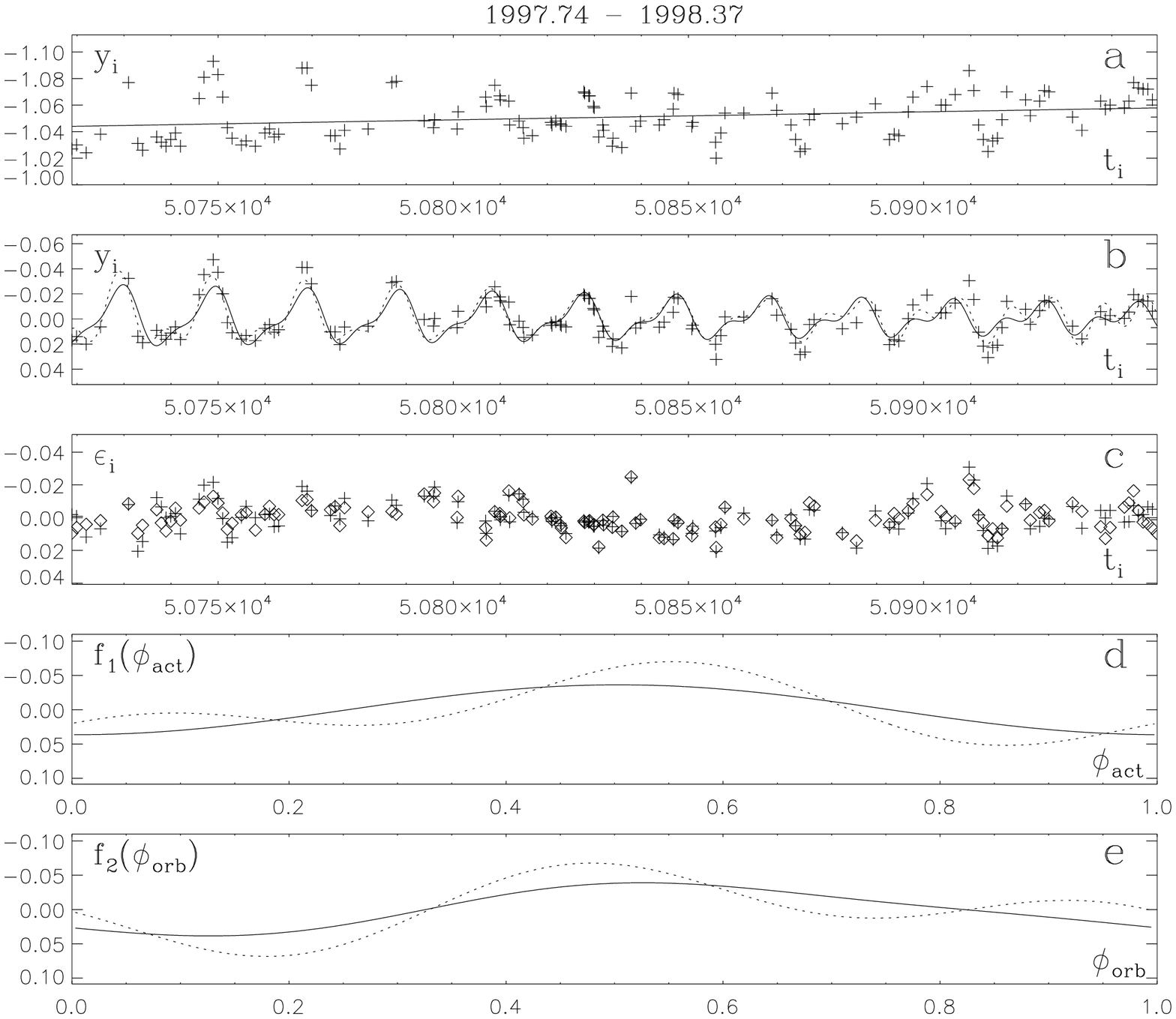}}
\caption{Another slice of $\sigma$~Gem photometry.
({\bf a}) Original photometry and a linear fit
({\bf b}) Original photometry minus linear fit
1st model 
(Eq. \ref{solution}: $K_1=1, K_2=2, M=M_1=M_2=0$, continuous line)
and 2nd model 
(Eq. \ref{solution}: $K_1=2, K_2=2, M=M_1=M_2=0$, dashed line)
({\bf c}) Residuals of 1st model (crosses) and 2nd model (diamonds)
({\bf d}) Nonstationary part $f_1(\phi_{\mathrm{act}})$
of 1st and 2nd model (continuous and dashed lines)
({\bf e}) Stationary part $f_2(\phi_{\mathrm{orb}})$
of 1st and 2nd model (continuous and dashed lines)
\label{slicetwo}}
\end{figure*}

\subsubsection{Argument 1: Orbital motion connection} 
\label{argumentone}

MLC of all data show the orbital phases {\it where} spots concentrate.
MLC of two separate samples of data, like those of the 1st and 2nd part,
reveal {\it where}  the largest spot distribution changes take place.
The \genflip ~model explains MLC of all fourteen CABS, especially the
connection between orbital phases and MLC primary and secondary minima.

\subsubsection{Argument 2: Mean and amplitude connection} \label{argumentwo}

MLC amplitude decreases when mean brightness increases 
(e.g. Fig. \ref{xxtri}),
and it increases when mean brightness decreases
(e.g. Fig. \ref{huvir}).
Only dark starspots can cause this effect.
The longitudinally evenly distributed spots cancel out in MLC, 
but longitudinally concentrated spots do not.
MLC reveals this mean and amplitude connection which
fits to the \genflip ~model.
For example,  Fig \ref{xxtri} indicates that
one side of \object{XX~Tri} is filled with spots that eventually disappear
at the brightness maximum having a low amplitude MLC.

\subsubsection{Argument 3: Single and double peaked MLC 
connection}
 \label{argumentthree}

Low and high MLC amplitudes are possible in S1f and S1b modes.
However, S12fb and S12bf modes can have only a low amplitude MLC, 
because the effects of S1 and S2 spots cancel out.
This is the reason, why the clearly double peaked MLC
have only low $A_{\mathrm{All}}$ values 
(Figs. \ref{v711tau}, \ref{eieri}, \ref{sigmagem},
\ref{hd89546}, \ref{v478lyr} and \ref{v1762cyg}).

\subsubsection{Argument 4: Ellipticity connection} \label{argumentfour}

The effects of ellipticity can be understood in the context
of the \genflip ~model (Fig. \ref{GFfig}: lowest line). 
Ellipticity can not cause two unequal MLC minima, while spots can.
Ellipticity amplifies MLC of S12fb and S12bf modes, but it
distorts MLC minima and maxima in S1f and S2b modes,
as well as MLC shape.
Ellipticity can weaken or strengthen MLC even 
in the same object, if its modes change.
For example, the aforementioned
MLC dips of \object{XX~Tri} (Fig. \ref{xxtri}g: $\phi_{\mathrm{orb}} \approx 0.25$ and 0.75)
and \object{V1762~Cyg} (Fig. \ref{v1149ori}eg: $\phi_{\mathrm{orb}} \approx 0.75$)
may be an ellipticity effect. 
In both cases, nearly all spots have faded away, 
and they do not mask the ellipticity effect. 

\subsubsection{Argument 5: \flip ~connection} 
\label{argumentfive}

The four types of \flip ~mode changes are

\begin{itemize}
\item[] Type I:   ${\mathrm{S1f}}     \leftrightarrow {\mathrm{S1b}} $
\item[] Type II:  $ {\mathrm{S12fb}}  \leftrightarrow {\mathrm{S12bf}}$
\item[] Type III: ${\mathrm{S1f}}     \leftrightarrow {\mathrm{S12bf}}$
\item[] Type IV:  ${\mathrm{S1b}}     \leftrightarrow {\mathrm{S12fb}}$
\end{itemize}

\noindent
The modes may also follow this order
\begin{itemize}
\item[]
${\mathrm{S1f}} \leftrightarrow {\mathrm{S12fb}} \leftrightarrow
{\mathrm{S12bf}} \leftrightarrow {\mathrm{S1b}}$.
\end{itemize}

For example, MLC of \object{EI~Eri}, \object{V478~Lyr} and 
\object{V1762~Cyg} show 
a Type II \flip ~(Figs. \ref{eieri}eg, \ref{v478lyr}eg and \ref{v1762cyg}eg).

\subsubsection{Some additional arguments}

{\it All} starspots of CABS are not circular, 
like in our \genflip ~model of Fig. \ref{genflip}.
They are not concentrated only on two active longitudes.
There are numerous other geometrical and physical phenomena
that can induce significant deviations from our simple model.
However, considering all these uncertainties,
\genflip ~model works surprisingly well.

The \flip ~phenomenon was originally
reported in the single G4 giant \object{FK~Com} \citep{Jet93}.
No unique period, like the orbital period of CABS, can be used to
compute MLC for single giants like \object{FK~Com},
or for single main sequence stars like \object{LQ~Hya} \citep[][K1~V]{Leh12}.
One solution might be to compute MLC for tested periods
within a short interval 
at both sides of the active longitude period.
The tested period with the highest MLC amplitude might
reveal similar results as we report here.

All our MLCs are not necessarily representative samples,
because we have not observed the full spot cycles of all CABS
(e.g. Fig. \ref{xxtri}a: \object{XX~Tri}).
It may also be that the \flip ~events are
connected to spot cycles \citep[e.g.][]{Ber05}.
A sliding window MLC within a time interval
shorter than $\Delta T$ may reveal such regularities.

Finally, we return back to \object{$\sigma$ Gem}, because the discovery
of its ellipticity \citep{Roe15} motivated the current study.
This star is a special case, where S1 and S2 nearly equally strong, 
and even small changes in spot areas can trigger a \flip ~events
``aided'' by the ellipticity effect.

\begin{figure*} 
\resizebox{\hsize}{!}{\includegraphics{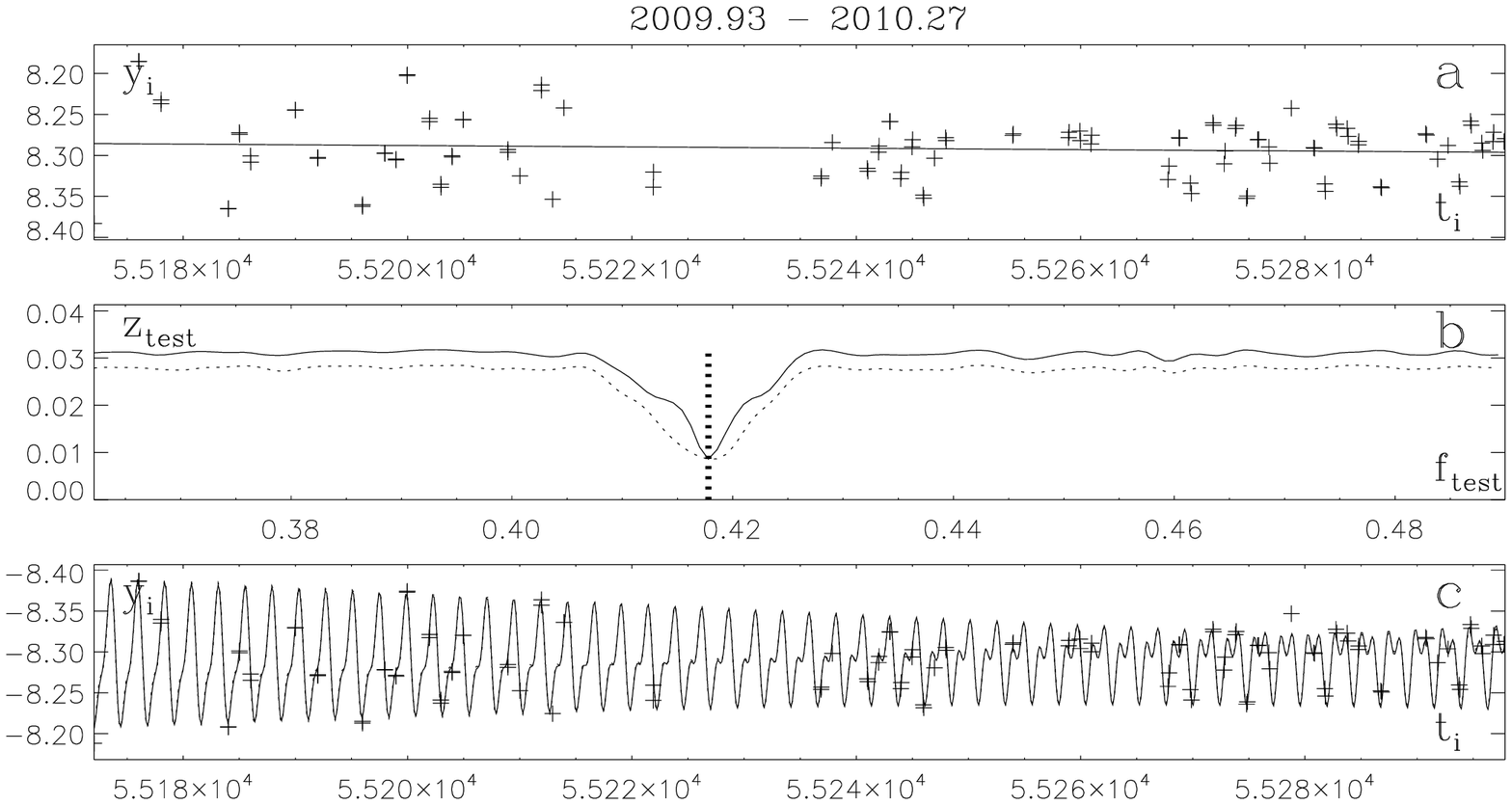}}
\caption{A slice of FK Com photometry.
({\bf a}) Original photometry and a linear fit
({\bf b}) $z_{\mathrm{test}}(f_{\mathrm{test}})$ periodograms for 1st model
(Eq. \ref{solution}: $K_1=1, K_2=2, M=M_1=M_2=0, P_{\mathrm{act}}=2.401151$:
continuous line),
and 2nd model
(Eq. \ref{solution}: $K_1=2, K_2=2, M=M_1=M_2=0, P_{\mathrm{act}}=2.401151$:
dashed line)
and
best period $P_{\mathrm{single}}=2.3935$ (vertical thick dashed line)
({\bf c}) Respective light curves of the above two models
with $P_{\mathrm{single}}=2.3935$: continuous and dashed line.}
\label{slicethree}
\end{figure*}

\subsection{Nonstationary part} \label{nonstationary}

The model for the binned MLC data (Eq. \ref{MLCmodel}) is stationary, 
because the active longitudes are locked to the synchronized
orbital motion and rotation frame with the period $P_{\mathrm{orb}}$. 
This model does not explain the regular migration of 
light curve minima (Figs. \ref{dmuma_long}--\ref{iipeg_long}: ``a'' panels).
Such migration has already been reported in binaries and single stars,
e.g. by \citet[][II~Peg]{Hac11}, \citet[][$\sigma$~Gem]{Kaj14} or
\citet[][FK Com]{Hac13}.
Our results in Figs. \ref{dmuma_long}--\ref{iipeg_long} 
show that the CABS light curves contain a nonstationary part
\begin{eqnarray}
f_1(t,\bar{\beta}_1)=
M_1 +
\sum_{k=1}^{K_2} 
A_k \cos ({{2 \pi k t} \over {P_{\mathrm{act}}}})
+
B_k \sin ({{2 \pi k t} \over {P_{\mathrm{act}}}})
\label{Eqnonstationary}
\end{eqnarray}
\noindent
where $\bar{\beta}_1=[M,A_1, ..., A_{K_1},
B_1, ..., B_{K_1}]$ are 
the free parameters and $K_2=1$ is probably sufficient,
because the $t_{\mathrm{CPS,min,1}}$ changes
have a linear connection to $\phi_{\mathrm{orb}}$
(Figs. \ref{dmuma_long}--\ref{iipeg_long}: panels ``a'').
This is the dark longitudinally dominating spot S3 
that rotates once around the surface of member A
during every $P_{\mathrm{cyc}}$ lap 
cycle (see Fig. \ref{Myway}: dark vertical line).
We use a vertical line to denote the longitude of this spot S3, 
because our figure is already quite crowded by spots S1 and S2.
If $f_1(t,\bar{\beta}_1)$ is double peaked, then
another nonstationary spot S4 rotates 180 degrees behind S3.
For obvious reasons, we have not tried to squeeze
spot S4 into Fig. \ref{Myway}.

MLC represent the second stationary part

\begin{eqnarray}
f_2(t,\bar{\beta}_2)
=
M_2 +
\sum_{k=1}^{K_1} 
C_k \cos ({{2 \pi k t} \over {P_{\mathrm{orb}}}})
+
D_k \sin ({{2 \pi k t} \over {P_{\mathrm{orb}}}})
\label{Eqstationary}
\end{eqnarray}
\noindent
where $\bar{\beta}=[M,C_1, ..., C_{K_1},
C_1, ..., C_{K_1}]$ are 
the free parameters, 
and $K_1=1$ or 2.
We compute the phases $\phi_{\mathrm{orb}}$
of this stationary part from the orbital period zero epoch
(Table \ref{resultstwo}: Ab epoch in Eq. \ref{thomasinvaiheet}).

Hence, the suitable CABS light curve model is
\begin{eqnarray}
f(t,\bar{\beta})=f_1(t,\bar{\beta}_1)+f_2(t,\bar{\beta}_2),
\label{solution}
\end{eqnarray}
\noindent
where the free parameters $M_1$ and $M_2$ are combined
to into one free parameter $M=M_1+M_2$.

This model is linear.
The nonstationary $f_1(t,\bar{\beta}_1)$ 
and stationary $f_2(t,\bar{\beta}_2)$ 
parts have a unique solution.
The interference between these two waves is
impossible to foresee without numerical modelling.
We will show how the model of Eq. \ref{solution} 
explains the four \flip ~types
of Sect. \ref{argumentfive}.

There are striking similarities in Figs. \ref{dmuma_long}-\ref{iipeg_long}.
The amplitude changes follow both $P_{\mathrm{cyc}}$ and $P_{\mathrm{orb}}$.
This reflects the interference of two ``waves''
(Eqs. \ref{Eqnonstationary} and \ref{Eqstationary}).
The \flip ~events tend to occur when the active longitudes migrate
through phases $\phi_{\mathrm{orb}}=0.25$ and 0.75. 
The relative strengths of the amplitudes of
the stationary part $f_2(t,\bar{\beta}_2)$ 
and the nonstationary part $f_1(t,\bar{\beta}_1)$
can be reversed at these particular phases.
This phenomenon is illustarated in the cases
when the minima first follow the horizontal
dotted lines and then begin to follow the the tilted continuous
or dashed lines, or vice versa 
(Figs. \ref{dmuma_long}-\ref{iipeg_long} :``a'' panels).
The two waves can either amplify each other or cancel out at these
same phases $\phi_{\mathrm{orb}}=0.25$ and 0.75.
Hence, there are numerous reasons for the large 
scatter of the amplitude values at these two phases. 
All these regularies allow us to 
identify the \flip ~events nearly unambiguously.

The sketches in Fig. \ref{Myway} illustrate one $P_{\mathrm{cyc}}$ 
lap cycle.
This model works like giant regular clock,
where the nonstationary wave rotates once around
member A during every lap cycle.
The locations of S1 and S2 spots are stable
in the synchronized orbital and rotational frame.
The sizes and/or temperatures of these spots change.
The dramatic events, like \flip ~phenomena, 
are more frequent when spot S3
crosses the longitudes of S1 and S2.
If the unseen secondary B member has spots,
its stationary spot configuration 
is most probably similar to that of A member. 
If this B secondary is a white dwarf, 
our model predicts its location with 
respect to S1 an S2 spots on member A.
It may now be easier to combine light curves to
surface imaging maps \citep[e.g.][]{Hac11,Lin13}, 
because our model gives the longitudes of 
spots S1, S2, S3 and S4 at any given epoch in time.

The results for \object{XX~Tri} reveal  that 
the free parameters of this model of Eq. \ref{solution}
depend on time,
e.g. the mean level of brightness $M=M(t)$.
The cycle for the mean of \object{XX~Tri} (see Fig. \ref{xxtri}a)
is clearly much longer than the $P_{\mathrm{cyc}}=7.8$ years lap
cycle in the light curve amplitude (Table \ref{resultsthree}).
This problem of temporally
changing free parameters can be solved by modelling the light curves
of Eq. \ref{solution}
during time intervals (i.e. windows) shorter 
than the whole time span of the data ($\Delta T$),
like with the CPS method \citep{Leh11}.
The information of the long--term
evolution of the magnetic fields of CABS is probably
coded into the free parameters of this model.
One advantage of our model is that 
the stationary and nonstationary parts of this magnetic
field can be uniquely 
separated and solved from the light curves.

Here is our short analysis recipe for CABS light curves: 
{\it 
Solve the active longitude period $P_{\mathrm{act}}$ from 
the minima epohcs ($t_{\mathrm{CPS,min,1}}$).
Use the known orbital period ($P_{\mathrm{orb}}$)
and apply the model (Eq. \ref{solution}) 
to the original photometry.}
We show two arbitrary slices of \object{$\sigma$ Gem} photometry
in Figs. \ref{sliceone} and \ref{slicetwo}.
The nonstationary and stationary orders
of the model for the 1st slice are $K_1=1$ and $K_2=2$ (Fig. \ref{sliceone}).
Note that our model gives
a unique solution also within gaps of data.
Subtraction of the same linear trend from the data at both sides
of the gap does not fully eliminate the mean level changes
(Figs. \ref{sliceone}ab), and the mean residuals
are therefore quite large $0.^{\mathrm{m}}012$ (Fig. \ref{sliceone}c).
The two models in Fig. \ref{slicetwo}b
have $K_1=1$ and $K_2=2$ (continuous line),
and $K_1=2$ and $K_2=2$ (dashed line).
Their mean residuals
are $0.^{\mathrm{m}}007$ (Fig. \ref{slicetwo}c: crosses)
and $0.^{\mathrm{m}}006$ (Fig. \ref{slicetwo}c: diamonds). 
These values are comparable to the accuracy of our data, $0.^{\mathrm{m}}006$.
Comparison of Figs. \ref{sliceone}de and \ref{slicetwo}de
shows that the nonstationary part, $f_1(\phi_{\mathrm{act}})$,
and stationary part, $f_2(\phi_{\mathrm{orb}})$, 
of this CABS are indeed changing.
The residuals of our model indicate that if the drift of short--lived
spots is present in $\sigma$ Gem, this effect is weak.
Although the presence of surface differential rotation
is not observed in the movement of starspots of \object{$\sigma$ Gem}, 
this does not prove that surface differential rotation is absent.
However, the stationary and nonstationary spots in \object{$\sigma$ Gem}
seem to defy differential rotation.

We give a slightly different recipe for single stars,
because they have no $P_{\mathrm{orb}}$ value:
{\it Solve the active longitude period $P_{\mathrm{act}}$ from the minima
epochs ($t_{\mathrm{CPS,min,1}}$). 
Rename $P_{\mathrm{orb}}$ to  $P_{\mathrm{single}}$ in Eq. \ref{Eqstationary}. 
Test a period interval $\pm 15$ \% at both sides of $P_{\mathrm{act}}$.
Fit our linear model (Eq. \ref{solution}) with each tested 
$P_{\mathrm{single}}$ value to the original photometry.
The $P_{\mathrm{single}}$ value
that gives the best fit to the data is the
rotation period of this single star.} 
A suitable test statistic is 
$z_{\mathrm{test}}=\sqrt{(1/n) \sum_{i=1}^{n} \epsilon_i^2}$,
where $\epsilon_i$ are the residuals of the model of Eq. \ref{solution}
with the known $P_{\mathrm{act}}$ 
value and the tested $P_{\mathrm{single}}$ value. 
This $z_{\mathrm{test}}$ is the mean of $|\epsilon_i|$ of each tested model.
We tested this method to
the single star \object{FK Com} using $P_{\mathrm{act}}=2.401151$ 
to a short slice of photometry from
\citet{Hac13}.
A linear fit to these data is shown in Fig. \ref{slicethree}a.
This linear trend was removed from the data before computing the
$z_{\mathrm{test}}$ periodograms.
We show the periodograms for two models 
(Eq. \ref{solution}: $K_1=1, K_2=2$)
and
(Eq. \ref{solution}: $K_1=2, K_2=2$)
in Fig. \ref{slicethree}b.
The best value was $P_{\mathrm{single}}=2.3935$.
The two solutions for the light curve of \object{FK Com}
overlap in Fig.  \ref{slicethree}c,
because their maximum difference is $0.^{\mathrm{m}}006$.
The mean residuals of these models are
$0.^{\mathrm{m}}009$ and $0.^{\mathrm{m}}008$.
The $P_{\mathrm{act}}$ period of \object{FK Com}
was determined from the light curve minima,
like in our CABS analysis. Hence, it represents the
nonstationary spot S3 detached from the  $P_{\mathrm{single}}$
rotation frame of \object{FK Com}. 
Refinement of $P_{\mathrm{act}}$ can {\it only}
be achieved from an analysis of new
$t_{\mathrm{CPS,min,1}}$ data.
Our $P_{\mathrm{single}}=2.3935$ value represents the stationary
part attached to the rotation frame of \object{FK Com}.
This value is certainly not final, because different slices 
of photometry from \citet{Hac13} gave different values.
A lot of data must be analysed using a much more careful subtraction 
of the changes of the mean level.
We are confident of that this
$P_{\mathrm{single}}$ value converges to an extremely accurate final value,
just like the orbital periods $P_{\mathrm{orb}}$ of a binary stars
(or the rotation periods of a single stars). 
Such a result is self evident, because both members
of all our CABS have  $P_{\mathrm{orb}}=P_{\mathrm{single}}$.
When the two stars in \object{FK Com} coalesced,
their $P_{\mathrm{orb}}$ became $P_{\mathrm{single}}$,
or equivalently, $P_{\mathrm{orb}}$ became meaningless.
We emphasize that the $P_{\mathrm{act}}$ value of \object{FK Com}
is already known, 
but more work is required to determine its $P_{\mathrm{single}}$ value.
This is one of the advantages of our model: the rotation periods of
single active stars are no longer based on the observed changing
$P_{\mathrm{phot}}$ values (Fig \ref{slicethree}b: only one clear minimum).

Let us say a few words about the light curve predictability
with Eq. \ref{solution}, for CABS and single active stars.
It is a least partly possible to predict the
observed light curves changes when 
the $P_{\mathrm{orb}}$ or $P_{\mathrm{single}}$, 
and $P_{\mathrm{act}}$ values are known.
The light curve minima follow the ``lattice''
$\phi_{\mathrm{orb}}=\phi_{\mathrm{single}}=0.25$ or 0.75,
and 
$\phi_{\mathrm{act}}=0.00$ or 0.5
(Figs. \ref{dmuma_long}-\ref{iipeg_long}: ``a''panels).
The difference $\Delta P_{\mathrm{orb/single-act}}= |P_{\mathrm{orb}}-P_{\mathrm{act}}|=
|P_{\mathrm{single}}-P_{\mathrm{act}}|$
determines the scale of observed light curve period changes. 
If the nonstationary part $f_1(t,\bar{\beta}_1)$ is weak, 
then $P_{\mathrm{act}}$ is absent, 
and the observed light curve period is close to $P_{\mathrm{orb}}$.
The opposite is true, when $f_2(t,\bar{\beta}_1)$ is weak.
If $\Delta P_{\mathrm{orb/single-act}}$ is small, the observed 
light curve period changes are small,
and vice versa.
The observed period changes reflect
this difference $\Delta P_{\mathrm{orb/single-act}}$ between two constant periods,
i.e. not surface differential rotation.
One period may dominate over the other one for a long period of time,
e.g. in \object{II~Peg} the stationary $P_{\mathrm{orb}}$ dominated
between the years 1987 and 1995, 
and then the nonstationary $P_{\mathrm{act}}$ began to dominate and
this domination has continued to the year 2015
(Fig. \ref{iipeg_long}a).
The case of \object{HK Lac} is very similar (Fig. \ref{hklac_long}a).
The $f_1(t,\bar{\beta}_1)$ and $f_2(t,\bar{\beta}_2)$ amplitudes
determine the maximum scale for the observed light curve amplitudes.
For example, perhaps the observed surface differential rotation 
in the light curves of slowly
rotating stars is not larger, 
but the ratio $\Delta P/P_{\mathrm{orb}}$ probably is.
As for another example, 
if  no active longitude is discovered,
then only the nonstationary part is present.
No swithces between the domination of  
$f_1(t,\bar{\beta}_1)$ and $f_2(t,\bar{\beta}_2)$
can occur.
This reduces the observed period variations
(The Sun and other inactive stars?). 
The results in \cite{Leh16} indicate that
the nonstationary part is absent in young inactive
solar--type stars, because no active longitudes
($P_{\mathrm{act}}$) are detected.

The periods $P_{\mathrm{orb}}$ and $P_{\mathrm{act}}$ of these CABS
are constant for long periods of time. 
The observed photometric periods of these stars appear to change, 
but this is probably caused by
the interference of two constant period waves. 
Does this then mean that there is no surface 
differential rotation in these CABS? 
It has been known for a long time that 
surface differential rotation in these rapidly
rotating stars is weak \citep[e.g.][]{Hal91A},
but no one has ever seriously claimed
that surface differential rotation is totally absent.
Surface differential rotation 
may be present in these CABS, but the locations of stationary S1 and S2 spots
seem stable, as well the movement of the nonstationary spots S3 and S4.
Could the driving mechanism of magnetic fields in these CABS
be the rotation of plasma of free charged particles
producing stationary and nonstationary magnetic field waves?
Spots S1--S4 are simply the signatures of these waves.
And where there is a wave, and another wave, there are probably numerous waves.
We have discovered only the two strongest waves in these CABS. 
Could the interference of stationary and nonstationary waves
sustain long-lived magnetic fields?
If these waves are stronger in rapidly rotating stars late type stars, 
their interference produces stronger spots. 
Could interference explain the differences between 
the magnetic fields of early and late type stars?
In early type stars, their radiative envelopes 
allow the waves of the magnetic field to remain stable.
In late type stars, convection complicates things.
Thus, quasiperiodic spot distribution 
changes are observed in late-type stars,
and stable spot distributions in early type stars?
A more sophisticated version of our model could be a sum 
of spherical functions describing the stationary and nonstationary 
magnetic field waves.
Perhaps the free parameters of this model
could be solved from observations.

This brings us back to the two currently known alternatives 
for magnetic fields in all spectral types 

\begin{itemize}

\item[1.] Rotation and convection (dynamo)

\item[2.] Fossil fields

\end{itemize}

We admittedly only speculate about a third alternative

\begin{itemize}

\item[3.] Rotation and interference

\end{itemize}

``Rotation'' is present only in two of these three alternatives.
Or is ``rotation'' present in all three alternatives?
Our expertise lies in time series analysis,
not in the modelling stellar magnetic fields.
Therefore, this interference hypothesis of ours is tentative.

Here is our simple analogy. 
The orbital and rotation periods of the Moon are equal. 
Due to this synchronization, we never see the dark side of Moon
when it is illuminated by the Sun.
Imagine that a giant large dark circular
screen would begin to rotate around the Moon
with a period of for example ten years ($P_{\mathrm{act}}$).
This is what we see in CABS and single stars.
The main differences are that the rotation period of Earth is not
synchronized with the orbital motion of the Moon,
and the ``spot'' on the Moon is not stuck to its surface,
but moves due to the illumination of the Sun.

\section{Conclusions}

We present a general model for the light curves 
of chromospherically active single and binary stars
(Eq. \ref{solution}).
This model explains the connection between orbital motion,
long-term starspot distribution changes, ellipticity and
\flip ~events in CABS.
The Ancient Egyptians discovered Algol's regular eclipses.
Some day this model of ours is perhaps referred to 
as something discovered by the Ancient Finns and Americans. 
The history of mankind pales
when compared to the millions or billions of years that
these CABS have already spent in synchronous orbital motion and rotation.
Two stars A and B are endlessly staring at each other's faces.
Spots S3 and S4 (if present) occasionally arrive to meet
spots S1 and S2 at every multiple epoch of half lap cycle $P_{\mathrm{cyc}}$.
Otherwise, this peaceful arrangement is nearly eternal.
This idea of ``nothing ever happens'' 
is something that truly works in astronomical time scales.
The geometry of the magnetic fields in these CABS
does in some ways begin to resemble 
the never changing geometry of the magnetic fields of Ap stars, 
the oblique rotator model \citep[e.g.][CQ~Uma]{Kok92}.
Are we observing interference waves of
``Oh Be A Fine Girl. Kiss Me, Right Now or Soon.''
 
\acknowledgements
This work has made use of the SIMBAD database at CDS, Strasbourg,
France and NASA's Astrophysics Data System (ADS) services.
The automated astronomy program at Tennessee State University
 has been supported by NASA, NSF, TSU and the State of Tennessee 
through the Centers of Excellence program.

\bibliographystyle{apj}                                   


\end{document}